\RequirePackage{fix-cm}
\documentclass[smallextended]{svjour3}       
\smartqed  
\usepackage{graphicx}
\usepackage{amssymb}
\usepackage{graphicx}
\usepackage{graphics, subfigure}
\usepackage{caption}
\usepackage{algorithmic}
\usepackage{algorithm}
\usepackage{url}
\usepackage{epstopdf}
\graphicspath{{./Figures/}}

\begin{document}

\title{On Discovering Co-Location Patterns in Datasets: A Case Study of Pollutants and Child Cancers
}
\titlerunning{On Discovering Co-Location Patterns in Datasets: A Case Study}        

\author{Jundong Li\footnote{The work is done when the author is at University of Alberta.} \and Aibek Adilmagambetov \and Mohomed Shazan Mohomed Jabbar \and Osmar R. Za\"{\i}ane  \and Alvaro Osornio-Vargas \and Osnat Wine
}

\authorrunning{Jundong Li et al.} 

\institute{Jundong Li \at
               Computer Science Engineering, Arizona State University, Tempe, Arizona, USA \\
               \email{jundong.li@asu.edu}
           \and
           Aibek Adilmagambetov \and Mohomed Shazan Mohomed Jabbar \and Osmar R. Za\"{\i}ane \at
              Department of Computing Science, University of Alberta, Edmonton, Alberta, Canada \\
              \email{\{adilmaga, mohomedj, zaiane\}@ualberta.ca}           
           \and
           Alvaro Osornio-Vargas \and Osnat Wine \at
              Department of Pediatrics, University of Alberta, Edmonton, Alberta, Canada\\
              \email{\{osornio, osnat\}@ualberta.ca}
}

\date{Received: Feb 12, 2014 / Accepted: March 28, 2016}

\maketitle

\begin{abstract}
We intend to identify relationships between cancer cases and pollutant emissions by proposing a novel co-location mining algorithm.  In this context, we specifically attempt to understand whether there is a relationship between  the location of a child diagnosed with cancer with any chemical combinations emitted from various facilities in that particular location. Co-location pattern mining intends to detect sets of spatial features frequently located in close proximity to each other. Most of the previous works in this domain are based on transaction-free apriori-like algorithms which are dependent on user-defined thresholds, and are designed for boolean data points. Due to the absence of a clear notion of transactions, it is nontrivial to use association rule mining techniques to tackle the co-location mining problem. Our proposed approach is focused on a grid based “transactionization” of the geographic space, and is designed to mine datasets with extended spatial objects. It is also capable of incorporating uncertainty of the existence of features to model real world scenarios more accurately. We eliminate the necessity of using a global threshold by introducing a statistical test to validate the significance of candidate co-location patterns and rules. Experiments on both synthetic and real datasets reveal that our algorithm can detect a considerable amount of statistically significant co-location patterns. In addition, we explain the data modelling framework which is used on real datasets of pollutants (PRTR/NPRI) and childhood cancer cases.

\end{abstract}

\section{Introduction}
\begin{figure}[!t]
\centering
\includegraphics[width=3in, height = 2.5in]{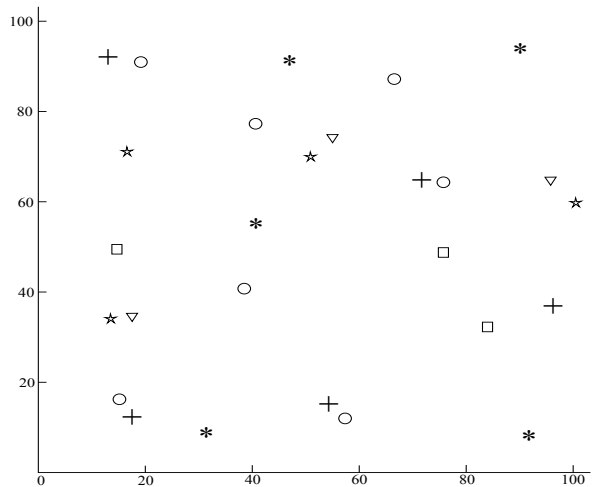}
\caption{A sample dataset with point spatial features. Instances of feature sets $\{+, \circ\}$ and $\{\star, \triangledown\}$ are often located close to each other}
\label{fig_sample}
\end{figure}

Co-location mining aims to discover patterns of spatial features often located close to each other, i.e. in geographic proximity. An example of a pattern is a co-location of symbiotic species of plants and animals depending on ecological conditions. Figure~\ref{fig_sample} illustrates a sample spatial dataset with point features. As it can be observed, instances of feature ``$+$" are often located close to instances of ``$\circ$". Similarly, objects of feature ``$\star$" are seen close to instances of ``$\triangledown$". The main purpose of co-location mining is to come up with a set of hypotheses based on data features and statistics that are potentially useful to domain experts, so that they can uncover possible patterns that are hidden in the data sets. The discovery of spatial co-location patterns may lead to useful knowledge in various applications. For instance, one might be interested in an animal species that lives close to certain types of landmarks such as rivers, meadows, forests, etc. Another example includes the co-location patterns detected between crime incidents and locations of various businesses, which can be useful for criminologists. Some of the application domains for co-location mining are biology, urban studies, health sciences, earth and atmospheric sciences, etc. Even though this task seems to be similar to association rule mining (ARM) which is used in knowledge discovery, the use and adaptation of ARM techniques are not trivial due to the fact that features are embedded into a geographic space and there is no clear notion of transactions. ARM consists of discovering rules that express associations between items in a database of transactions. The associations are based on a notion of frequency bounded by a given threshold, such as the minimum frequency.

\begin{figure}[!t]
\centering
\includegraphics[width=3in]{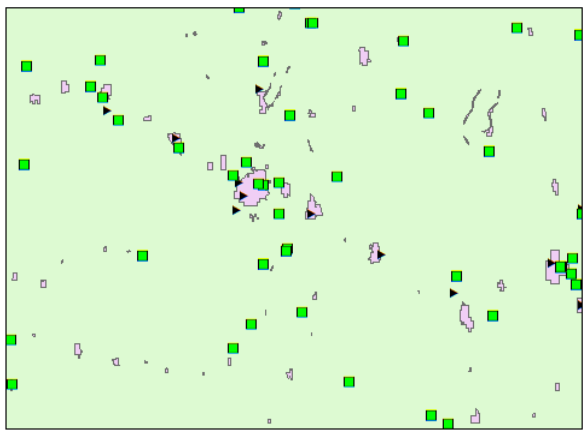}
\caption{Part of the dataset from the province of Alberta: rectangles - pollutant points, triangles - cancer cases, polygons - urban municipalities}
\label{fig_map}
\end{figure}

An application which motivates the work of this paper is the detection of possible spatial associations of different chemicals and cases of childhood cancer. Cancer, a multifactorial class of diseases, characterized by uncontrolled growth of abnormal cells, their invasion into other tissues, and metastasis, is one of the leading causes of adults' death in both the developed and developing world~\cite{armstrong1975environmental,boffetta2003contribution}. Although some individuals are genetically predisposed to cancer, most cases of cancer are suspected to be linked to environmental factors such as air pollutants, radiation, various infections, tobacco, and alcohol. However, causes of childhood cancer are difficult to determine, partially because of the fact that children's cancer cases are relatively rare and the levels of exposure to environmental factors are difficult to evaluate. Our collaborative research efforts with the Faculty of Medicine at the University of Alberta attempt to identify associations between cancer cases and known chemical emissions from the industry. Some of these chemicals are proven to be carcinogens while others are not known to cause cancer on their own. It is yet to be discovered whether certain combinations of chemicals can be associated with higher rates of cancer. 
Moreover, even if chemicals in such combinations with a potential threat are not emitted by the same industry, atmospheric conditions can contribute to the mixture. Given these concerns, we deploy our model to detect co-location patterns on a spatial dataset which contains information on chemical emission points, amounts of released chemicals and the locations of childhood cancer cases if they were first diagnosed in the two Canadian provinces, Alberta and Manitoba, where the data are collected from. Figure~\ref{fig_map} displays a part of the dataset with rectangles representing pollutant emission points, triangles for cancer cases, and polygons for urban municipalities. To mine such a rich dataset, first we build a modeling framework which handles the data to represent the real world conditions as accurate as possible while taking various factors which affect distribution of chemicals into account. This underlying modeling framework helps the co-location mining algorithms we develop to detect more accurate patterns. While we are not intending to find causalities, the goal of our study is to identify potentially interesting spatial associations in order to state hypotheses and further investigate relationships between childhood cancer and specific combinations of chemicals.

Most of the existing approaches to the co-location mining problem~\cite{Shekhar2001,Huang2006,Yoo2006,Xiong2004} deploy a framework which requires a user-defined minimum prevalence threshold. Without prior knowledge it could be difficult to choose a proper threshold. Furthermore, spatial features often have various frequencies in datasets, and one global threshold might lead to the omission of some co-location patterns and rules with rare events or the detection of meaningless patterns. Another limitation of most of the existing algorithms is that they work with point spatial features and a single neighborhood distance threshold, whereas in reality there are datasets which, in addition to point instances, also have lines and polygons (e.g., a road network map). Moreover, the information in some datasets can be uncertain: the presence of a feature in the region could depend on different factors, thus associating it with an existential probability. For example, a pollutant released from a facility distributes according to the climatic factors in its area, and the probability of detecting the chemical in a region close to the emission point is higher than in remote regions.

In this paper, we address these limitations in existing approaches by proposing a new framework which combines ideas from co-location mining, frequent pattern mining and association rule mining. Our proposed framework uses statistical tests to determine the significance of co-location patterns and rules. A co-location pattern or rule is considered as significant if it has a surprisingly high level of prevalence in comparison with randomized datasets which are generated under the null hypothesis that the spatial features are independent from each other. The uncertainty of the information is modeled as a dependent relationship based on the distance from the spatial object. We also paid attention to the computational complexity of the proposed algorithms. In this paper, we discuss some of the filtering techniques we used to increase the efficiency of the algorithms.

In one of our previous papers \cite{adilmagambetov2013discovering} we outlined some of the core ideas of the framework proposed in this paper and provided some preliminary results. In this paper, we extend the above work in few different aspects. First, we provide a concrete description of our complete co-location mining framework by properly defining the Grid Transactionization algorithm and including it in the workflow of the mining process. Furthermore we include details on how to work with uncertainty in Spatial datasets. We also tested our proposed framework on a completely new, real dataset for the province of Manitoba, Canada, which allowed us to analyze the robustness and effectiveness of our approach while also helping to derive new co-location rules. We also include results from further experiments conducted on synthetic datasets to help validate the results. Results from another variation of our proposed framework are also included, which ignores the uncertainty attribute of the spatial dataset. The results from this ``certain method" assist in further exploring the extensibility of the proposed framework. We also implemented a baseline co-location rule miner and compared the rules mined by that with the rules mined by our approach. This has provided more evidence that our algorithm is not only capable of deriving highly prevalent rules but also it can detect statistically significant rules which may not be highly prevalent. Furthermore, we provide a closer analysis on the computational complexity of the proposed algorithms, as well as on the choice of a proper grid granularity measure.

The remainder of the paper is organized as follows. An overview of the related work is given in Section 2. The proposed framework and its outline are described in Section 3. Section 4 describes the challenges and the modeling framework used to mine the co-location patterns between pollutants and childhood cancer cases. The experiments are presented in Section 5, followed by conclusions in Section 6.

\section{Related Work}
In recent years spatial data mining has gained significant attention due to the abundance of data with spatial and temporal attributes. In particular spatial data mining is the process of extracting interesting and useful patterns in geographic datasets. The technological advances in data storage and the widespread use of GPS technologies, remote sensing devices, and location-based services have created large amounts of spatial data. The spatial data processing and analysis is useful in a wide range of applications such as population analysis, social sciences, environmental sciences, and business applications.

In contrast to classical data mining, spatial data mining has some specific features. In classical data mining, it is assumed that data objects are independent from each other, whereas in spatial datasets, objects situated close to each other tend to be more similar, and have the same characteristics than objects located at a great distance. Autocorrelation is another important feature which can be observed in spatial datasets that helps to explain the gradual change of temperature with precipitation levels. One of the major difficulty in dealing with spatial data is the relatively higher complexity of its data object types and their relations. For instance, there are not only points but also lines and polygons in spatial databases. Furthermore, the relationship between objects are implicit, such as intersection, containment and enclosure.

Various types of methods and approaches are used to analyze such spatial dasets. Insights from traditional data mining techniques have been useful in developing these methods. Some of the tasks in spatial data analysis include spatial clustering, co-location mining, spatial trend detection, outlier detection and spatial classification~\cite{Ester2001,Shekhar2010}. We discuss some of the significant previous works in this context under two major categories:1) Co-location Mining as a category of spatial data mining techniques, and 2) Frequent Pattern Mining as a general data mining technique which can provide insights in co-location pattern detection.
\subsection{Co-location Mining}
Co-location mining is one of the tasks of spatial data mining that can be divided into two classes of methods: spatial statistics approaches, and spatial data mining approaches.
\subsubsection{Spatial Statistics Approaches}
Spatial Statistics Approaches deploy statistical techniques such as cross K-functions with Monte-Carlo simulations~\cite{Cressie1991}, mean nearest-neighbor distance, and spatial regression models~\cite{Chou1997} to evaluate and find co-location patterns between two features. Disadvantages of these approaches are expensive computation time and the difficulty in applying them to patterns consisting of more than two spatial features.
\subsubsection{Spatial Data Mining Approaches}
Spatial Data Mining Approaches could be categorized into transaction-based methods (i.e. works with transactions), and spatial join-based methods (i.e. use spatial joins of instance tables or feature layers).

Transaction-based approaches work by creating transactions over the space and using association rule mining like algorithms on these transactions~\cite{Agrawal1994,Koperski1995,Morimoto2001}. One of these methods, a reference centric model~\cite{Koperski1995}, creates transactions around a reference feature specified by the user. Each set of spatial features that form neighbourhood relationships with an instance of the reference feature is considered as a transaction. However, not all applications have a clearly defined reference feature. For example, in urban studies, features could be schools, fire stations, hospitals, etc., and there is no single specific feature of interest. Another approach, the window-centric model~\cite{Shekhar2001}, divides the space into cells, and considers instances in each cell as a transaction. The model can consider all possible windows as transactions or use spatially disjoint cells. However, a major drawback of the model is that some instance sets are divided by the boundaries of cells. Therefore some of the spatial relationship information is lost. On the other hand maximal cliques (i.e. maximal sets of instances which are pair-wise neighbors) in spatial data are also proposed to be used as transactions~\cite{Naymat2008,Kim2011}. However this approach does not preserve the information regarding the relative distance of the objects in cliques as long as they are considered neighbours.

Spatial join-based approaches work with spatial data directly. They include cluster-and-overlay methods and instance-join methods. In the cluster-and-overlay approach, clustering is used to mine associations. For example, concentrations of objects in layers are found in order to search for possible causal features~\cite{Estivill-Castro1998}. In another work~\cite{Estivill-Castro2001}, layers of points and area data are constructed for each spatial feature based on clusters of data instances or boundaries of those clusters. The above proposes two algorithmic approaches for cluster association rule mining: vertical-view approach and horizontal-view approach. In the former, after the clusters in the layers are found, the layers are segmented into a finite number of cells. Then, a relational table is constructed where an element is equal to one, if the corresponding cell satisfies an event in a layer, and the element is zero otherwise. Afterwards an association rule mining algorithm is applied to that table. The second approach evaluates intersections of clustered layers. A clustered spatial association rule defined in the above work is of the form $X\rightarrow Y(CS\%,CC\%)$, where $X$ and $Y$ are the sets of layers, $CS\%$ is the clustered support (i.e. the ratio of the area that satisfies both $X$ and $Y$ to the total area of the study region), and $CC\%$ is the clustered confidence (i.e. the percentage of cluster areas of $X$ that intersect with clusters of $Y$). However, these approaches might be highly sensitive to the choice of clustering methods. In addition, this cluster based approach assumes that features are clustered, even though spatial features may not form explicit clusters. The other spatial join-based approach, the instance-join algorithm, is similar to classical association rule mining. One of the first proposed co-location pattern mining frameworks of this type~\cite{Shekhar2001,Huang2004} is based on the neighbourhood relations and the participation index concepts.

Following the previous cluster-overlay methods to detect co-location patterns ~\cite{Estivill-Castro2001}, another approach~\cite{Huang2006dens} uses clustering to detect co-location patterns as follows. For two spatial features $f_{1}$ and $f_{2}$, if the density of objects of the feature $f_{1}$ in proximity of objects of the feature $f_{2}$ is higher than the overall density of objects of $f_{1}$, then the feature $f_{1}$ is considered to be co-located with the feature $f_{2}$, i.e. their objects tend to be situated close to each other. Similar to some of the previous approaches, this algorithm also suffers from a limitation where it is based on an assumption that spatial instances of a feature are situated close to each other and form clusters which may not be the case in some real-world applications.

The basic concepts in the co-location mining frameworks are analogous to the concepts in traditional association rule mining. As an input, the framework takes a set of spatial features and a set of instances, where each instance is a vector that contains information on the instance ID, the feature type of the instance, and the location of the instance. As an output, the method returns a set of co-location rules, where a co-location rule is of the form $C_{1}\rightarrow C_{2}(PI,cp)$, where $C_{1}$ and $C_{2}$ are co-location patterns, $PI$ is the prevalence measure (i.e. the participation index), and $cp$ is the conditional probability. The participation index $PI(C)$ of a co-location pattern $C$ is defined as:

\begin{equation}
PI(C) = min_{f_{i}\in C}\{pr(C,f_{i})\}
\end{equation}
where $pr\{C,f_{i}\}$ is the participation ratio of a feature $f_{i}$ in a co-location $C$ and it is computed as:

\begin{equation}
pr(C,f_{i}) = \frac{\textnormal{\# distinct instances of } f_{i} \textnormal{ in instances of } C}{\textnormal{\# instances of } f_{i}}
\end{equation}

A co-location pattern is considered prevalent, or interesting, if for each feature of the pattern, at least $PI\%$ instances of that feature form a clique with the instances of all the other features of the pattern according to the neighbourhood relationship. Similar to association rule mining, only frequent $(k-1)$ patterns are used for the $k$ candidate generation process. A co-location rule $C_{1}\rightarrow C_{2}$ is considered prevalent if its conditional probability is higher than a threshold. The conditional probability $cp(C_{1}\rightarrow C_{2})$ is defined as:
\begin{equation}
cp(C_{1}\rightarrow C_{2}) = \frac{\textnormal{\# distinct instances of }C_{1} \textnormal{ in instances of }C_{1} \rightarrow C_{2}}{\textnormal{\# instances of }C_{1}}
\end{equation}

In the approach mentioned above, it is assumed that spatial features occur with similar levels of frequency. Therefore, if a dataset contains rare spatial features, co-locations involving these rare events will be pruned by a prevalence threshold because more frequent features dominate rare ones, and no pattern with a rare event can become prevalent. For example, a rare disease will not be captured in co-location patterns due to the fact that its causes are more frequent in the database. To solve this limitation, Huang et al.~\cite{Huang2006} continue their previous work by introducing an algorithm that finds co-location patterns with rare features. Instead of the participation index threshold, the authors propose to use the maximal participation ratio threshold. Briefly, a co-location pattern is considered prevalent if $maxPR\%$ instances of at least one of the features in the pattern are co-located with instances of all the other features, where $maxPR$ is the maximal participation ratio:
\begin{equation}
maxPR(C) = max_{f_{i}\in C}\{pr(C,f_{i})\}
\end{equation}
It is not well explained how the algorithm deals with noisy features. They identify noisy features as the features that have a relatively insignificant number of instances in the given spatial dataset than the other features, and those instances are concentrated only on few co-location patterns. In this case it is highly probable that every co-location pattern with these features will be considered prevalent because of their high participation ratio irrespective of their insignificant representation in the whole dataset. Both previously mentioned methods use computationally expensive instance joins to identify instances of co-location patterns, and their running time grows fast as the number of instances and sizes of candidate patterns increase.

Yoo et al.~\cite{Yoo2004} propose a partial-join approach for mining co-location patterns. A study space is partitioned into square cells with the side length equal to a neighbourhood distance threshold. A set of spatial instances in a cell form a clique. Join operations are required to identify neighbourhood relationships divided by boundaries of cells. Even though this approach reduces the computation time, it still requires a large amount of spatial joins.

The joinless algorithm~\cite{Yoo2006} is a follow-up work to the partial-join approach. It further decreases computation time of constructing neighbourhood relationships. The main idea is to find star neighbourhoods instead of calculating pairwise distances between all the instances in a dataset. The neighbourhood relationship is materialized in the form of a table where for each instance, all its neighbors are listed. Then, in order to ensure that pattern instances form cliques, an instance-lookup scheme is used to filter co-location instances. In addition, three filtering steps are used to find a set of prevalent co-location patterns. The authors prove that their algorithm finds a complete and correct set of co-location patterns and rules. The experiments on synthetic and real datasets show that the joinless approach has better performance in terms of the running time than the join-based algorithm.

Based on their work, Xiao et al.~\cite{Xiao2008} improve the running time by dividing spatial objects into partitions and detecting neighboring instances in dense regions first. The algorithm finds instances in dense regions and maintains an upper bound on a prevalence measure for a candidate pattern. If the upper bound becomes less than a threshold, the method decides that it is a false candidate and stops identifying its instances in less dense regions.

Several other studies have extended the basic co-location mining framework to work with more complex spatial objects. For example, Xiong et al.~\cite{Xiong2004} propose a framework for detecting patterns in datasets with extended objects. Extended objects are objects that are not limited to spatial points but also include lines and polygons. This framework also uses the notion of buffers, which can be defined as zones of specified distances created around spatial objects. The size and the shape of these buffer zones might depend on the types of the spatial objects. In the proposed model, candidate patterns are pruned by a coverage ratio threshold. In other words, if an area covered by the features of a candidate pattern is greater than a predefined threshold, this pattern is considered prevalent. In order to minimize the usage of geographic information systems (GIS) overlay methods, a coarse-level mining step is used. At this level, minimum buffer bounding boxes of spatial objects are considered by the algorithm instead of true buffer shapes. Then, patterns that have coarse level coverage ratio higher than the threshold are evaluated using actual buffers. Compared to previous models, this approach takes into account the shapes of spatial objects and their distribution in space rather than using a single neighbourhood distance for varying types of features. Expensive GIS overlays are used in this method and a filtering technique is proposed in order to improve its performance.

The approaches mentioned above use thresholds on interestingness measures, which causes meaningless patterns to be considered as significant with a low threshold, and a high threshold may prune interesting rare patterns. Instead of a threshold based approach, Barua and Sander~\cite{Barua2011} use a statistical test to mine frequent co-location patterns. The participation index of a pattern in observed data is calculated as in previous studies. Then, for each co-location pattern the authors compute a probability $p$ of seeing the same or a greater value of the prevalence measure under a null hypothesis model. A co-location is considered significant if $p\leq \alpha$, where $\alpha$ is the level of significance. However, the statistical significance is not a monotonic property and it cannot be used to prune insignificant co-location rules as apriori-like algorithms. Thus in their work, they limit the size of the co-location pattern/rule and test each possible candidate pattern/rule to see if it passes the statistical test. To generate more general co-location rules without rule size constraint, Li et al.~\cite{Li2014discovering} propose a new co-location pattern mining framework by exploiting the property of statistical significance. The results of co-location rules are hard to evaluate even for domain experts, they also propose to use a classifier to help evaluate the results of co-location rules

\subsection{Frequent Pattern Mining}
The co-location mining problem is similar to the canonical data mining problem: association rule mining. The most classical example of association rule mining is discovering sets of goods that are often bought together. The concepts of association rule mining and co-location mining are compared in Table~\ref{table_ARM}.

\begin{table}[!t]
\caption{A comparison of association rule mining and co-location mining~\cite{Huang2006}.}
\label{table_ARM}
\centering
\begin{tabular}{|l|l|}
\hline
{\bf Association rule mining} & {\bf Co-location mining}\\
\hline
Item & Spatial feature\\
\hline
Itemset & Spatial feature set\\
\hline
Frequent pattern & Co-location pattern\\
\hline
Support \& Confidence & Interestingness measures\\
\hline
Transactional database & Spatial database\\
\hline
\end{tabular}
\end{table}

Considering the similarity between frequent pattern mining and co-location pattern mining, we discuss the classical frequent pattern mining problem in this subsection. The concepts of frequent pattern and association rule mining were first introduced by Agrawal et al.~\cite{Agrawal1993}. Various approaches to these problems have been proposed over the past two decades. Apriori~\cite{Agrawal1994} is the first and one of the most-known algorithms used for frequent itemset mining. This approach is designed to work on transactional data and consists of a bottom-up candidate generation process where $k$-size candidate itemsets are generated from frequent $(k-1)$-itemsets and tested against the database to obtain frequent $k$-itemsets. This process is repeated until no more candidate patterns can be generated. The correctness of the algorithm is based on the downward closure, or apriori, property, which states that if an itemset is frequent, then all its subsets are also frequent. In other words, an itemset cannot be frequent if one of its subsets is infrequent.

The ARM problem is defined as follows. Let $I = \{i_1, i_2, ..., i_m\}$ be a set of $m$ items and $T = \{t_1, t_2, ..., t_n\}$ be a set of $n$ transactions where a transaction $t$ is a subset of items in $I$. For an itemset $X \subseteq I$, the support of $X$ is defined as the ratio of transactions in $T$ that contain instances of $X$. An itemset is considered frequent if its support is higher than a user-specified minimum support threshold. An association rule is a rule of the form $X \to Y$, where $X \subseteq I$, $Y \subseteq I$, and $X \cap Y = \emptyset$. The confidence of a rule $X \to Y$ is the support of $X \cup Y$ divided by the support of $X$.

\begin{equation}
conf(X \to Y)= \frac{sup(X \cup Y)}{sup(X)}.
\end{equation}

\subsubsection{Frequent Pattern Mining with Uncertain Data}
The algorithms and approaches mentioned above are constructed to work with data where the presence of items in transactions is certain (i.e. definite). For example, market basket datasets are certain and precise (i.e. we know for sure the items were purchased). However, in some applications, data may be incomplete or may have errors. For instance, sensor reading records might include some erroneous data due to various internal and external factors such as sensor failures or extreme weather conditions. Uncertainty can be expressed in terms of existential probabilities; each item of a transaction is followed by a probability of its existence in this transaction. An example transactional dataset is shown in Table~\ref{table_transactions}.

\begin{table}[!t]
\caption{An example of a probabilistic transactional dataset.}
\label{table_transactions}
\centering
\begin{tabular}{|c|l|}
\hline
ID & Transaction\\
\hline
1 & A(0.7); B(1.0); C(0.2)\\
\hline
2 & A(0.9); D(0.5); E(0.4); F(0.8)\\
\hline
3 & B(0.3); D(1.0); G(0.7)\\
\hline
4 & A(0.1); B(0.6); C(0.7); E(0.2); G(0.4)\\
\hline
5 & C(0.5); D(0.2); E(0.8)\\
\hline
6 & B(0.6); C(0.3); E(1.0); F(0.4)\\
\hline
\end{tabular}
\end{table}

Most studies use a notion of expected support~\cite{Chui2007,Chui2008} to mine frequent patterns from uncertain databases. The expected support $E(s(I))$ of an itemset $I$ is defined as the sum of expected probabilities of the presence of $I$ in each of the transactions in a database. A probability $p(I, T)$ of presence of $I$ in a transaction $T$ is a product of corresponding probabilities of items in the transaction. An itemset is considered significant if its expected support exceeds a $minsup$ (i.e. minimum support) threshold.

Several approaches to frequent pattern mining problem with uncertain data have been studied by Aggarwal et al.~\cite{Aggarwal2009}. These approaches are extended from existing classical frequent itemset mining methods and can be divided into two categories: candidate generate-and-test algorithms (an extension of the Apriori algorithm) and pattern growth algorithms (extensions of FP-growth~\cite{Han2004} and H-Mine~\cite{Pei2001}). According to this study, while FP-growth is efficient and scalable in the deterministic case, its extension to the uncertain case behaves differently due to the challenges associated with uncertain data. UH-Mine, an extension of H-Mine, is reported to provide the best trade-offs in terms of running time and memory usage.

Bernecker et al.~\cite{Bernecker2009} proposed a PFIM (Probabilistic Frequent Itemset Mining) framework which is based on the possible world method. Instead of the expected support, PFIM uses the frequentness probability as a significance measure. By using a dynamic computation method, the algorithm is reported to run in $O(|T| minsup)$, where $|T|$ is the number of transactions and $minsup$ is a user-defined threshold. Without it, the approach runs in exponential time. However, the algorithm requires the $minsup$ threshold to be defined, and it is nontrivial to apply the statistical test to the frequentness probability.

\section{Proposed Algorithm}
As we discussed in the previous section, various approaches to the co-location mining problem have been proposed during the past decade. Most of them were focused on improving the performance of existing frameworks that had some limitations. Several studies have addressed some of these issues one at a time, but not all of them at the same time. Yet, these algorithms are unable to be used in some real-world applications, such as our motivating question of exploring whether co-locations of cancer cases and sets of released chemicals exist. We discuss some of these issues and limitations under three categories as follows.

First, the usage of the prevalence measure threshold for the detection of interesting co-location patterns and rules is a main limitation factor in many co-location mining algorithms. In spatial datasets, features usually have a varying number of instances; they could be extremely rare or be present in abundance. Therefore, one threshold for the participation index (or any other significance measure) cannot capture all meaningful patterns, while other patterns could be reported as significant even if their relation is caused by autocorrelation or other factors. In addition, most of the existing algorithms use a candidate generation process which forms $(k+1)$-size candidates patterns or rules only from significant $k$-size patterns. However, a set of features could be interesting even if some of its subsets are not significant. For example, two chemicals may not be correlated with a particular disease on their own, but could be correlated when in combination. In our approach we do not follow such a $(k+1)$-size candidate generation process from $(k)$-size patterns. Instead we use a statistical test to replace the prevalence measure threshold and to identify statistically significant co-location patterns from a given set of candidate patterns. Barua and Sander~\cite{Barua2011} first proposed the usage of statistical tests to find significant co-location patterns. In such a statistical framework, the pattern is considered as significant if the probability of seeing a similar or greater value of the prevalence measures in $N$ artificial datasets than in the observed or original dataset is less than $\alpha$ (the significance level) under the null hypothesis that there is no spatial dependency among features of the pattern. In this approach each candidate pattern is evaluated separately for their statistical significant rather than applying a single prevalence threshold for all of them.

Second, most of the existing co-location mining approaches use a single distance threshold to identify neighbourhood relationships among spatial objects. However, in some applications this might oversimplify the real situation. For instance, in zoological research, various species have different habitat ranges: birds (especially, birds of prey) might interact with other species in large distances, while subterranean animals are limited in their movements. Therefore, the usage of a single distance threshold might lead to inaccurate results. Furthermore, most of these existing approaches are designed to work with point datasets. However, other types of objects may exist in spatial datasets such as lines (i.e. roads and communication networks) and polygons (i.e. polluted regions, or areas that had no precipitation for some period or were exposed to other climatic factors). Even though some of the frameworks for extended spatial objects~\cite{Xiong2004} deal with lines and polygons, they also use a global threshold value as the prevalence measure. Furthermore, this framework cannot deal with the uncertainty in datasets (as explained in the following paragraph).

Third, in some applications, information in datasets can be uncertain; data may be incomplete or may have errors. For example, distribution of a chemical released from a chimney in a polluted region is not uniform. Areas closer to an emission point are generally exposed to higher pollutions than places far away from the release point. Another example is climatic data collected by sensors which may have errors in their readings. Uncertainty can be expressed in terms of existential probabilities where each item of the transaction is followed by the probability of its existence in this transaction. Although uncertainty in datasets has been researched for the frequent itemset mining problem, to the best of our knowledge there is no such work done for spatial data and the co-location pattern mining problem.

In this paper, we propose a new framework that addresses the aforementioned limitations. It uses a grid-based ``transactionization" approach (i.e. creating transactions from a dataset). The statistical test is performed on the derived set of transactions to get significant co-location rules or patterns. In the following section we explain the methods and the algorithms in our framework.

\subsection{Algorithm Design}
\label{ad}
The objective of this work is to detect significant patterns in a given spatial dataset that have a prevalence measure value higher than the expected value. Such spatial datasets may contain points and extended spatial objects such as lines or polygons. We use buffers, which are zones of specified distances created around spatial objects, to define the area affected by the instances in the dataset. For example, a buffer defined around a chemical emission point shows the area polluted by a released chemical. The size and the shape of these buffers may depend on the types of spatial instances depending on various factors, which may vary for different applications. In addition, the likelihood or the probability of the presence of the corresponding feature in the zone covered by the spatial object and its buffer is not uniform, and may depend on factors such as the distance from the object. Considering the above factors, we propose a new transaction-based co-location pattern mining algorithm that is suitable for extended spatial objects and uncertain data.
\subsubsection{Grid-based Transactionization}
Recall that previous transaction-based methods have some limitations. A window-centric model cuts off neighbourhood relations of instances located close to each other, but in different partitions. A reference-centric model may get duplicate counts of spatial instances. In addition, it is nontrivial to generalize this approach to applications with no reference feature, as in the case of our motivating application. Due to these limitations in previous transactionization models we propose a new grid-based transactionization method. Our grid-based transactionization procedure is outlined in Algorithm~\ref{alg_transaction}: \textit{GetTransactions(S)}.

\begin{figure*}[!t]
\centering
\begin{minipage}{1.3in}
\subfigure[A sample spatial dataset with point feature instances and their buffers]
{\includegraphics[width=1.3in]{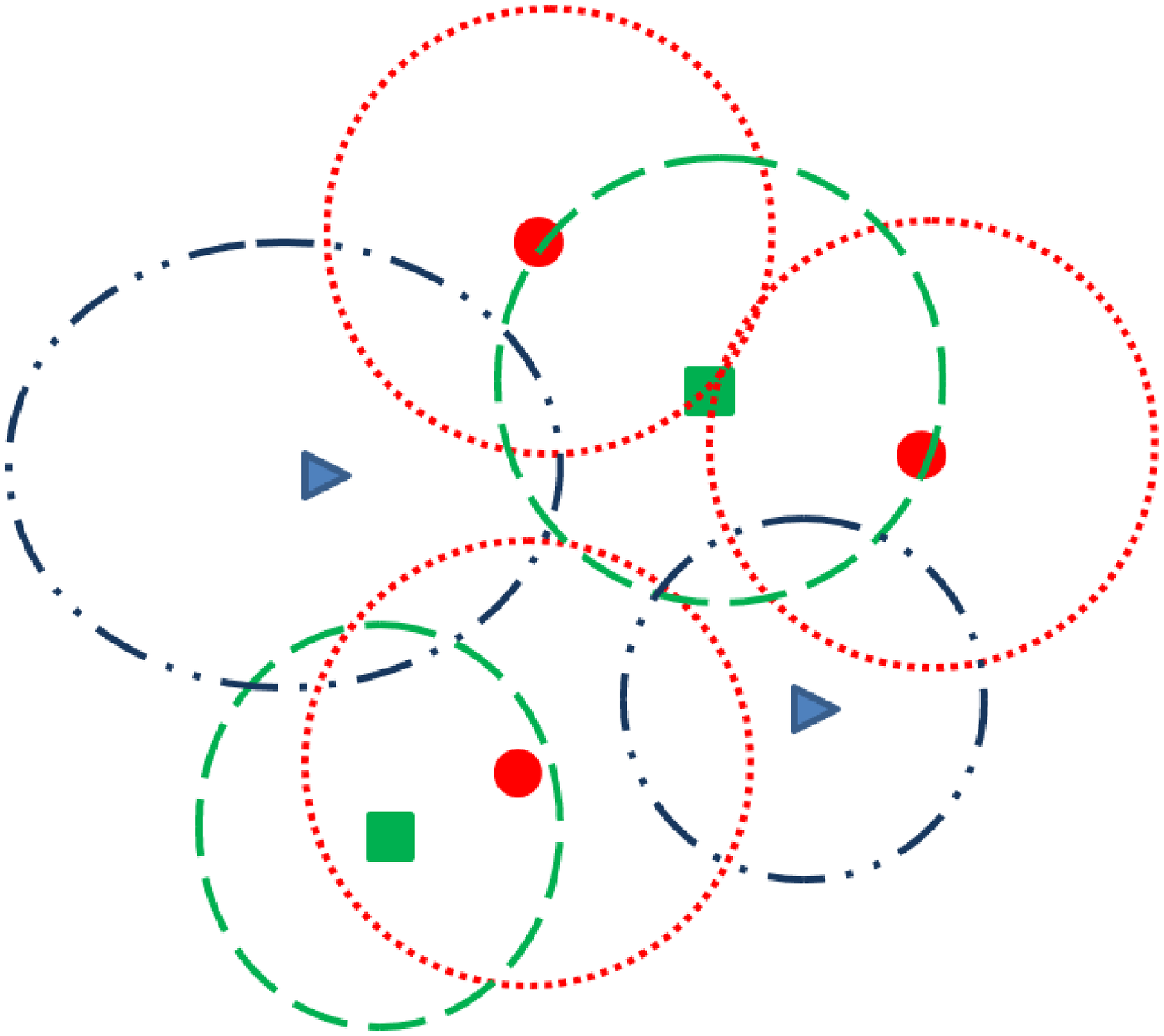}
\label{fig_grid_a}}
\end{minipage}\hspace{0.1in}
\begin{minipage}{1.3in}
\subfigure[A grid imposed over the space]
{\includegraphics[width=1.3in]{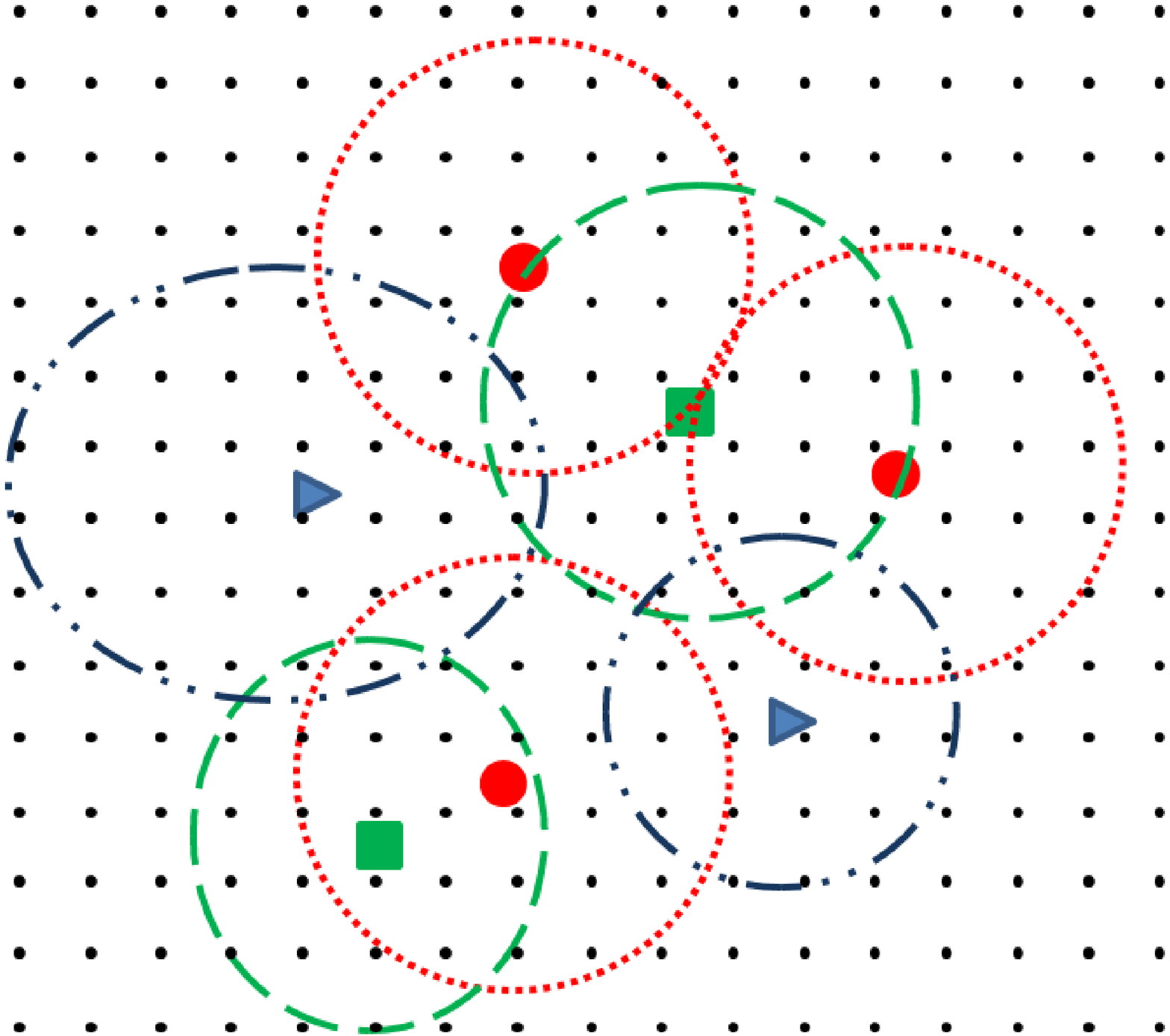}
\label{fig_grid_b}}
\end{minipage}\hspace{0.1in}
\begin{minipage}{1.3in}
\subfigure[Grid points which intersect with buffers are used to create transactions]
{\includegraphics[width=1.3in]{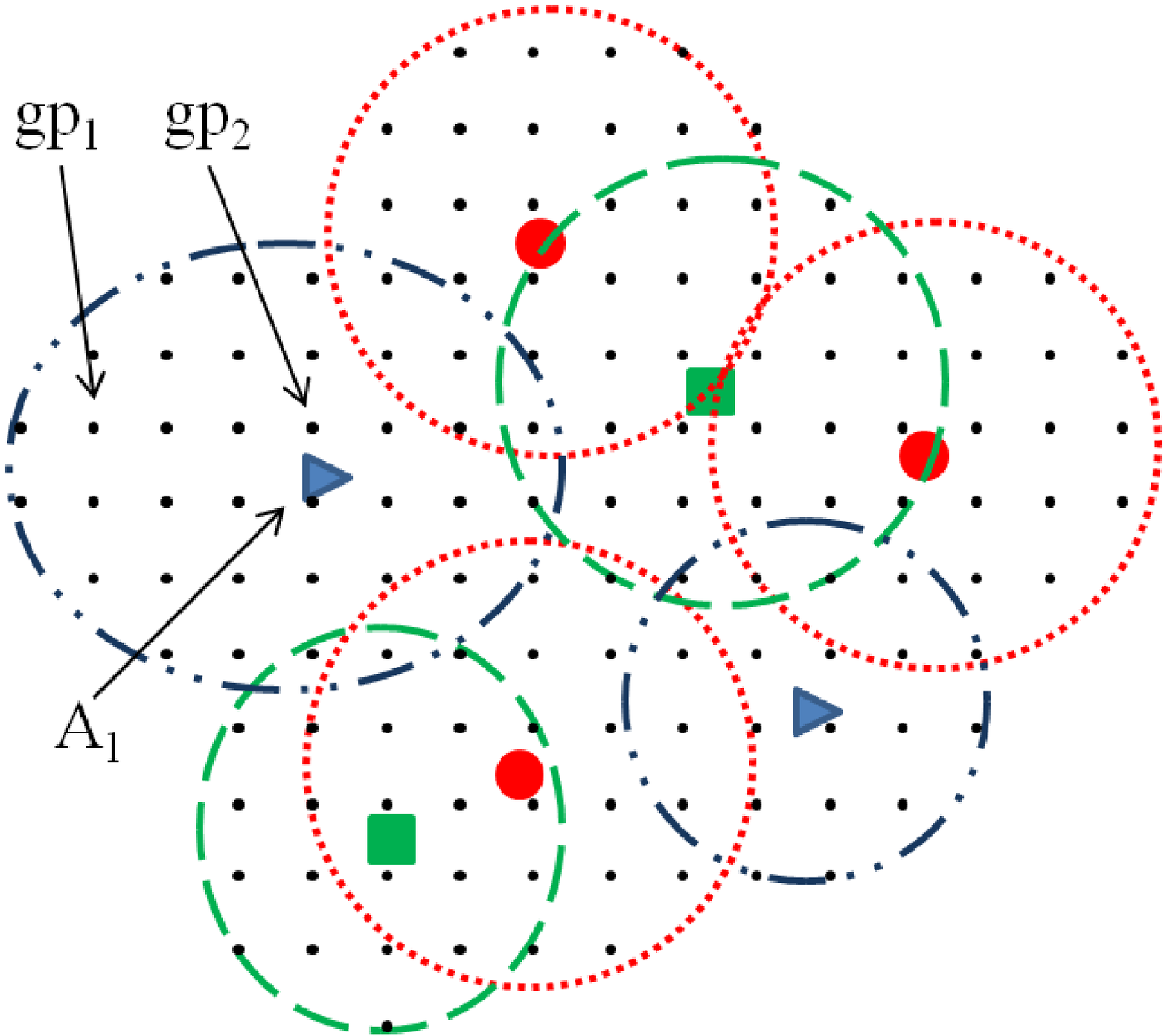}
\label{fig_grid_c}}
\end{minipage}
\centering
\caption{Transactionization step}
\label{fig_grid}
\end{figure*}

Given a spatial dataset \textit{S}, in its first step, Algorithm~\ref{alg_transaction} obtains the set of grid points by imposing a grid with a suitable granularity over the geographic space covering the spatial points in \textit{S}. Each point in this grid can be seen as a representation of a respective part of the corresponding geographic space. Once the grid points are obtained, then Algorithm~\ref{alg_transaction} constructs buffer zones around spatial objects in \textit{S}. In the dataset of our motivating application we have two types of spatial objects: 1) Childhood cancer cases and, 2) chemical emission points. Defining buffer zones for these two types of spatial objects are performed in a different manner. A childhood cancer case spatial object represents the location of a patient who is diagnosed with cancer. For such objects we define a fixed buffer which is a circular region around the source point with a fixed radius. This region defines the mobility range of the patient. On the other hand, when defining buffer zones for chemical emission points we consider factors such as the amount of chemicals emitted, wind speed and the direction. This is further discussed in Section~\ref{mf}: Modelling Framework. There we explain how we morph an original circular buffer region based on the emission quantity for chemical emission points, to an elliptical region which more realistically represents the chemical dispersion based on other factors such as the wind. In the next step of the algorithm the constructed grid is imposed over the dataset \textit{S}. Figure~\ref{fig_grid_a} illustrates an example dataset with buffers around spatial point instances, and a grid is laid over it in Figure~\ref{fig_grid_b}. Similarly, buffers can also be created around linear and polygonal spatial objects. In a two-dimensional space, grid points represent a square regular grid. Due to the spheroid shape of the Earth, a grid used for real-world applications becomes irregular. However, with a careful choice of a grid granularity this fact should not considerably affect the accuracy of the results.

A grid point may intersect with one or several spatial objects and their buffers. A transaction is defined as a set of features corresponding to these objects. Hence each grid point can be considered as a potential candidate to obtain a transaction. Let us assume that a sensor capable of detecting various features is placed at each grid point. A set of features detected by each sensor can be seen as a transaction. However, sensor readings are not fully reliable; they are uncertain and can be affected by extreme environmental conditions, sensors' hardware, durability, and other factors. For example, it is possible that a sensor detects a pollutant only if certain amount of it is present in the sensor's environment. In addition, a likelihood of a presence of a feature in a region covered by an object and its buffer is not uniform. Alternatively, since we work with estimates and not with sensors and sensor collected data, we can use in our model the notion of concentration of features. While the fading concentration is not a probability, it can be used to show the feasibility of our model using uncertainty. Intuitively, a feature is more likely to be detected in buffer parts which are closer to a feature point than in parts that are farther away from it. Furthermore, spatial datasets can be noisy and contain errors; locations of instances and their presence can be uncertain.

In order to take into account these uncertainties, the probability of a feature being present in a transaction is also stored when computing the transaction associated with each grid point in the steps six and seven of Algorithm~\ref{alg_transaction}. One of the ways to model uncertainty when transforming a spatial dataset into a set of transactions is to use a distance from a spatial object to a grid point (our method of estimating the existential probability of a feature is explained in the following section). For example, the grid point $gp_2$ in Figure~\ref{fig_grid_c} is located closer to the point $A_1$ than point $gp_1$; we can assume that $p(A,gp_2)>p(A,gp_1)$. An example transaction set is shown in Table~\ref{table_transactions}. When a grid point intersects with several instances of the same feature or their buffers, the highest existential probability is taken as the probability of detecting this feature at the given grid point. Moreover, the granularity of the grid, or the distance between points of the grid, should be carefully chosen for each application, and it may depend on an average size of a region covered by a spatial object and its buffer. A large distance between grid points may negatively affect the accuracy of the results because small feature regions and their overlaps may get a different number of intersecting grid points depending on a grid imposition. On the other hand, when the distance between grid points is too short, the number of derived transactions increases, and the following computation to discover the significance of candidate co-location patterns or rules might become prohibitively expensive, especially when the number of candidates is large.

\begin{algorithm}[!t]
\caption{{\it GetTransactions(S)}: Transactionization step.}
\label{alg_transaction}
\begin{algorithmic}[1]
\STATE $T = \emptyset$: set of transactions
\STATE $G$: set of grid points
\STATE Build buffer zones around spatial objects of $S$
\STATE Impose a grid $G$ over the dataset $S$
\FORALL {point $g \in G$}
\STATE $t$ = get a set of features which instances contain $g$ with corresponding existential probabilities
\STATE $T = T \cup t$
\ENDFOR
\RETURN $T$
\end{algorithmic}
\end{algorithm}

\subsubsection{Co-location Pattern/Rule Mining Algorithm}
Once the transactions are derived using Algorithm~\ref{alg_transaction}, we perform co-location pattern mining using Algorithm~\ref{alg_main}. As the first step after the transactionization, given the set of derived transactions $T$, and a set of spatial features $F$, Algorithm~\ref{alg_main} computes a prevalence measure value (i.e. the level of interestingness) for all the candidate co-location patterns or rules. Depending on the objective this prevalence measure could be different. In some applications, experts look for sets of features that are often co-located with each other, but not necessarily in a cause-effect relationship. In this case, which is analogous to the frequent pattern mining, the expected support $ExpSup(P)$ might be used to define a level of interestingness of a pattern $P$. In frequent pattern mining with certain data, the support of a pattern is counted deterministically as the number of transactions containing all features of the pattern. However, in the case of uncertain data, transactions are probabilistic, and therefore, the support is counted in the expected value, which is defined as follows.
\begin{definition}
The expected support $ExpSup(P)$ of a pattern $P$ is defined as the sum of probabilities of the presence of $P$ in each of the transactions $t$ in the uncertain database:
\begin{equation}ExpSup(P)=\sum \limits_{t \in T} p(P,t).\end{equation}
\end{definition}

The probability of a presence of a pattern $P$ in a transaction $t$ can be computed as follows:
\begin{definition}
The probability $p(P,t)$ of the pattern $P$ occurring in a transaction $t$ is the product of corresponding feature instance probabilities:
\begin{equation}p(P,t)= \prod \limits_{f \in P}p(f,t).\end{equation}
\end{definition}

For some other applications, researchers intend to discover a predefined set of rules. For example, for a dataset of disease outbreaks and possible cause factors, a typical co-location rule is of the form $C\rightarrow D$, where $C$ is the set of cause features and $D$ is the disease feature. For this scenario, the expected confidence $ExpConf(X \to Y)$ can be used as the prevalence measure of a co-location rule $(X \to Y)$, where $X \subseteq F$, $Y \subseteq F$, and $X \cap Y = \emptyset$. The definition of $ExpConf(X\rightarrow Y)$ is as follows.
\begin{definition}
The expected confidence $ExpConf(X \to Y)$ of a rule $X \to Y$ is defined as:
\begin{equation}ExpConf(X \to Y)=\frac{ExpSup(X \cup Y)}{ExpSup(X)}.\end{equation}
\end{definition}

Hence, based on the objective whether to mine rules or patterns, our algorithm uses either $ExpSup(P)$ or $ExpConf(X \to Y)$ as the prevalence measure in Algorithm~\ref{alg_main}. For example, Algorithm~\ref{alg_main} shows the pseudocode of our approach in a case where co-location patterns are mined and the expected support is used as the prevalence measure.

In the steps of Algorithm~\ref{alg_main} discussed above, we build buffers around each instance and then perform grid transactionization to obtain a set of transactions from the spatial dataset, and compute a prevalence measure value for each of the candidate co-location pattern or rule using the derived transactions. Now our goal is to discover a set of significant patterns and rules. As discussed above, the usage of one threshold on the prevalence measure may result in the discovery of wrong patterns and the omission of interesting ones. Moreover, only co-location patterns or rules with surprising levels of the prevalence measure should be considered as significant. In other words, it is unlikely that instances of features in a significant pattern are located close to each other only by chance according to a predefined significance level threshold.

Thus, in order to avoid the possibility that instances of features are co-located together by chance, we need to do some randomization tests; the instances of features are considered to be independent in each randomized test. Therefore, in this step, we utilize the statistical test to help estimate the likelihood of seeing the same, or a greater, level of the prevalence measure under a null hypothesis that the features of a pattern or rule are spatially independent from each other, i.e. the randomization test. The definition of significance is as follows:
\begin{definition}
A co-location pattern $P$ is considered significant at level $\alpha$, if the probability $p$ of detecting an expected support higher or similar to the $ExpSup_{obs}(P)$ (i.e. expected support in the original dataset) in a dataset complying with the null hypothesis is not greater than $\alpha$.
\end{definition}

The same logic can be applied to a case when significant co-location rules are mined:
\begin{definition}
A co-location rule $R$ is considered significant at level $\alpha$, if the probability $p$ of detecting an expected confidence higher or similar to the $ExpConf_{obs}(R)$ (i.e. expected confidence in the original dataset) in a dataset complying with the null hypothesis is not greater than $\alpha$.
\end{definition}

Let us suppose that we are mining for co-location rules. In this case, Algorithm~\ref{alg_main} uses the expected confidence $ExpConf$ as the prevalence measure. Let $ExpConf_{obs}(X \to Y)$ denote the expected confidence of a co-location rule $X \to Y$ in a real dataset, and $ExpConf_{rand}(X \to Y)$ denote the expected confidence of a rule $X \to Y$ in a randomized dataset which is generated under the null hypothesis. The expected confidence of the co-location rule in each of $R$ randomized datasets is calculated in order to estimate the probability $p$. Having the number of simulations $R$, the value of $p$ is computed as:
\begin{equation}\label{eq_pi_value}p=\frac{R_{\ge ExpConf_{obs}}+1}{R+1},\end{equation}
where $R_{\ge ExpConf_{obs}}$ is the number of simulations in which $ExpConf_{rand}(X \to Y) \ge ExpConf_{obs}(X \to Y)$. The observed dataset is added to both the numerator and the denominator.

If the $p$-value is less or equal to a predefined level of significance $\alpha$, the null hypothesis is rejected. Therefore, it is unlikely that the features of the rule are spatially independent; thus, they are not situated close to each other only by chance. Hence, the co-location rule $X \to Y$ is considered significant at level $\alpha$. The above explanation can also illustrate a process of detecting co-location patterns. The difference is that, instead of the expected confidence, the expected support $ExpSup$ is used as the prevalence measure.

In an incremental manner, Algorithm~\ref{alg_main} computes the $p$-value described above for each candidate pattern or rule as depicted in the iterative loop from step 10 to 22. In order to estimate this probability $p$, a set of randomized datasets (i.e. $RD[1...R]$) is generated under the null hypothesis. Each randomized dataset has the same number of instances of each feature as in the original dataset. In addition, the distribution of instances of each feature in a randomized dataset should be similar to its distribution in the original data. For instance, disease cases should be placed within populated areas. Obviously, the random placement of disease cases all over the study region can lead to invalid results, especially in the case when most of the region is unpopulated. Another example can be found in biology. Some animal species may have various requirements to their habitats such as a location close to water reservoirs or presence of certain types of vegetation. This observation needs to be taken into account in a randomized dataset generation process. When computing the $p$-values and identifying significant rules or patterns, rather than going through the full iteration, we perform a candidate filtering which would help reduce the computation time. In the following discussion we explain how this computation of $p$ values and detecting significant rules works with randomized datasets and the candidate filtering technique.

\begin{algorithm}[!t]
\caption{Mining significant co-location patterns.}
\label{alg_main}
\begin{algorithmic}[1]
\REQUIRE Spatial dataset $D$.\\Level of significance $\alpha$.\\Number of simulation runs $R$.\\Set of randomized spatial datasets $RD[1..R]$.
\ENSURE Set of significant co-location patterns $P$
\STATE $T$: set of transactions
\STATE $CP$: set of candidate patterns
\STATE $T=GetTransactions(D)$
\FORALL {$cp \in CP$}
\STATE $cp.ExpSup_{obs} = Compute ExpSup(cp,T)$
\IF {$cp.ExpSup_{obs} == 0$}
\STATE $CP = CP \backslash cp$
\ENDIF
\ENDFOR
\FOR {$i = 1 \to R$}
\STATE $T=GetTransactions(RD_i)$
\FORALL {$cp \in CP$}
\STATE $cp.ExpSup_{sim}[i] = Compute ExpSup(cp,T)$
\IF {$cp.ExpSup_{sim}[i] \ge cp.ExpSup_{obs}$}
\STATE $cp.R_{\ge ExpSup_{obs}} = cp.R_{\ge ExpSup_{obs}}+1$
\STATE $cp.\alpha = \frac {cp.R_{\ge ExpSup_{obs}}+1}{R+1}$
\IF {$cp.\alpha > \alpha$}
\STATE $CP = CP \backslash cp$
\ENDIF
\ENDIF
\ENDFOR
\ENDFOR
\STATE $P = CP$
\RETURN $P$
\end{algorithmic}
\end{algorithm}

The calculation of the $p$-value is repeated for all candidate co-location patterns or rules. The number of candidates grows exponentially with the number of spatial features in the dataset. In addition, the accuracy of the $p$-value depends on the number of simulation runs. Therefore, the more randomized datasets are checked, the more accurate are the results. These two factors may lead to an enormous amount of computation. However, the support of a co-location decreases as the size of a candidate pattern or rule increases, because fewer transactions contain all its features. Therefore, researchers may put a threshold on the support or the maximal size of a candidate in order to analyze only the patterns and rules that are backed by a meaningful number of transactions. That is one important constraint we use when applying the algorithm in real datasets. In addition, we use the following filtering techniques to prune candidate patterns or rules that are definitely not significant from the p-value computation in Algorithm~\ref{alg_main}.

\begin{enumerate}
\item After the calculation of the prevalence measure for candidate patterns in the real spatial dataset, some of the patterns may have a prevalence measure value equal to zero. It means that combinations of features of these patterns do not exist in the dataset. Obviously, these patterns cannot be statistically significant and they can be excluded from the set of candidate patterns (see steps 6-8 in Algorithm~\ref{alg_main}). In some other applications, a different, low-value threshold on a prevalence measure can be used in order to get significant patterns and rules with a certain level of interestingness. In this case, candidate patterns and rules with a prevalence value lower than this threshold can also be pruned and excluded from further computations.

\item During the calculation of the $p$-value for candidate patterns for which an observed prevalence is higher than zero (i.e in the main iterative loop from step 10-22 in Algorithm~\ref{alg_main}), some of the candidate patterns might show a $p$-value that has already exceeded the level $\alpha$. For example, let us assume that the number of simulation runs is 99 and $\alpha=0.05$ (which is most commonly used $p$-value to claim significance). If after ten simulation runs, the prevalence measure of a pattern $P$ is greater than the observed prevalence in 5 randomized datasets, pattern $P$ already surpassed the threshold ($(5+1)/(99+1) > 0.05$). Therefore, it definitely cannot be significant and can be excluded from the following 89 checks (see steps 16-19 in Algorithm~\ref{alg_main}). Thus, the computation time is greatly reduced. With this filter, after the last simulation run the remaining set of candidates contains only significant patterns or rules allowing Algorithm~\ref{alg_main} to return this remaining set as the statistically significant rules set.
\end{enumerate}

\subsection{Computational Complexity}
\label{cc}
The computational time complexity or cost of the proposed algorithm is very important to understand the effectiveness, efficiency and the practicality of it. Let's first analyze the time cost of Algorithm~\ref{alg_transaction}. A closer look at the algorithm reveals that there are four major operations which influence the computational cost of the GetTransactions(S) procedure (i.e. Algorithm~\ref{alg_transaction}). Let's define \textit{n} as the number of chemical emission points and \textit{m} as the number of cancer cases.
\begin{enumerate}
\item Obtain the set of grid points G (step 2 of Algorithm~\ref{alg_transaction}): This operation first selects maximum/minimum latitudes and longitudes from the recorded cancer cases and chemical emission points which causes an O(n+m) computational complexity. Then given the granularity of the grid (distance between two grid points) the operation would compute the grid points by iterating from the maximum latitude to minimum latitude and in each of these iterations it would also iterate from the maximum longitude to the minimum longitude. This would give a computational complexity of $O(k_{1}k_{2})$, where $k_1$ is the number of divisions from the maximum latitude to the minimum latitude and $k_2$ is the number of divisions from the maximum longitude to the minimum longitude. Let's also introduce $k_3$, where $k_3= k_1k_2$, which is the number of grid points.
\item Build the buffer zones around the spatial objects (step 3 of Algorithm~\ref{alg_transaction}) : This operation defines a buffer or a range which is a circle or an elliptical area around a particular spatial point or an object. This buffer construction takes a constant time and can be defined as $k_4$. Computational cost incurred by this operation can be represented as O($k_4$(n+m)).
\item Impose the grid G over the dataset S (step 4 of the Algorithm~\ref{alg_transaction}): This operation requires to join each of the record in the input data (size of (n+m)) with each of the grid points, which leads to the run time complexity of O($k_3$(n+m)).
\item Compute the transactions (step 5-8 of the Algorithm~\ref{alg_transaction}) : This operation groups the temporary transactions obtained by joining input data records with the grid points in the previous step. This causes a run time cost in the order of O($k_3$(n+m)). Then each of the categories is processed to compute the count of the feature instances or to identify the maximum existential probability of a particular feature. This would incur a constant cost \textit{c} on average. This would give a run time complexity in the order of O(c$k_3$).
\end{enumerate}
Based on the above, the overall run time complexity of the GetTransaction(S) algorithm can be represented as, $O(k_3)$+O($k_4$(n+m))+O(2$k_3$(n+m))+O(c$k_3$). If simplified by fixing all the other input except the original dataset which contains chemical emission points and cancer cases, we obtain, O(n+m) for Algorithm~\ref{alg_transaction}. However, the grid granularity can play a major role in deciding the value of the constants $k_1,k_2$ and $k_3$ giving the possibility of them to grow to a large number in the order of hundreds of thousands. Other constants such as $k_4$ and c can be safely considered as having an insignificant effect.
\par
When analyzing the computational complexity of Algorithm~\ref{alg_main} it can be recognized that there are three major operations. The computational complexity of these operations would determine the overall time cost of Algorithm~\ref{alg_main}.
\begin{enumerate}
\item GetTransactions(D) (i.e. Algorithm~\ref{alg_transaction}): The computational complexity of this method is in the order of O(n+m) as it is discussed previously.
\item Compute expected support of the candidate patterns: Part of this procedure may be recognized as constructing a candidate pattern set. Based on the number of features and the size of the itemset this could gain a computational complexity in the order of O($|f|^d$), where $|f|$ is the size of the feature set $F$ and the $d$ is the size of the largest itemset. However since the number of features and the size of the largest itemset can be considered as fixed constants, in reality, this complexity does not affect the time cost hugely. On the other hand computing the expected support can cost a time complexity of O($k_3|f|^d$).
\item Execute simulation runs and compute expected support with randomized datasets: This would have a run time complexity in the order of O(R$k_3|f|^d$), where R is the number of Randomized datasets and when the other costs to compare the expected support in the actual and randomized datasets, and to compute and compare the p-values are considered constant and insignificant.
\end{enumerate}
Based on this the overall run time complexity of Algorithm~\ref{alg_main} can be represented as, $O(n+m)$+O($k_3|f|^d$)+O(R$k_3|f|^d$). This emphasizes that the original number of inputs plays a minor role in determining the overall computational complexity of the Algorithm, whereas the number of Randomized datasets/ simulation times, R, number of grid points, $k_3$, (i.e. indirectly the grid granularity), and the largest itemset size ($d$) play a major role in determining the overall computational complexity of Algorithm~\ref{alg_main}. However, since each randomization or simulation is done independently, Algorithm~\ref{alg_main} can be done in an ``embarrassingly parallel" procedure with $R$ parallel nodes.
\subsection{Advantages of the Proposed Algorithm}
As we have shown in our previous discussions, by combining techniques of co-location mining and frequent pattern mining, we address the limitations in previous co-location mining approaches. Using our framework has the following advantages over others:
\begin{itemize}
\item The main advantage of our framework is that the only parameter it requires is the $p-value$ (i.e. level of significance). Other than that our framework does not need thresholds on prevalence measures. The statistical test replaces a usage of one global threshold for a prevalence measure of candidate co-location patterns or rules. Only meaningful patterns are reported as significant. These patterns have the prevalence measure higher than an expected value under a null hypothesis that features of a pattern are independent from each other. Sometimes researchers do not need patterns or rules with very low support values even if they are significant. In this case a threshold on support can be used. However, it should have a relatively low value in comparison with other approaches, so it does not exclude meaningful patterns or rules.

\item While a neighbourhood distance threshold used in many co-location algorithms is set to one value for all spatial features, our model can deal with varying buffer sizes. A buffer size may depend on the types of features or on attributes of individual spatial objects. So, the algorithm can be used for applications where features differ from each other with regard to the effect to the environment around them, e.g., plant and animal species. Moreover, the model can be further extended to take into account not only point instances but also other types of spatial objects such as lines and polygons which are present in many spatial datasets.

\begin{figure}[!t]
\centering
\includegraphics[width=3.5in]{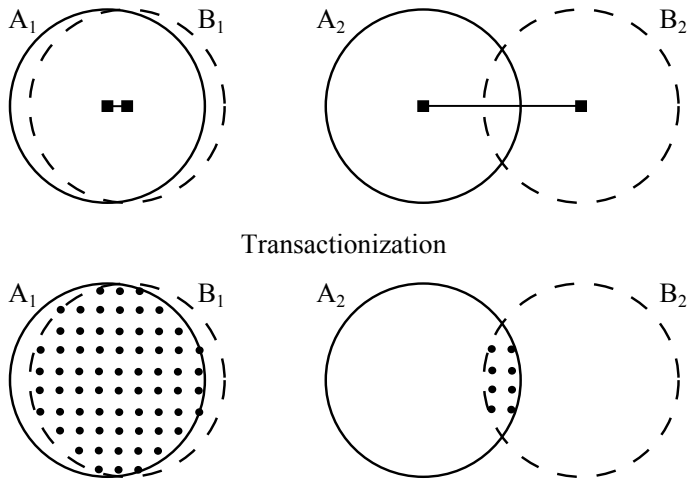}
\caption{Neighboring objects $A_1 - B_1$ and $A_2 - B_2$. In the transactionization step, the intersection of $A_1$ and $B_1$ receives more transactions (black dots) than the pair $A_2$ and $B_2$}
\label{fig_prox_remote}
\end{figure}

\item In most previous algorithms, two or more objects are considered to have a neighbourhood relationship, if they are located at a distance not farther than a distance threshold. However, these approaches do not take into account spatial information and context: how close or far the objects are situated from each other. Figure~\ref{fig_prox_remote} illustrates an example of two pairs of neighboring spatial objects with corresponding buffer zones. Both pairs, $A_{1} - B_{1}$ and $A_{2} - B_{2}$, are neighbors and treated similarly by most co-location approaches, even though $A_1$ and $B_1$ are closer to each other than $A_2$ to $B_2$. Being located at a closer distance, the instances of the former pair are more likely to be related than the instances of the latter pair. By using buffer zones around spatial instances and transactions that are created from grid points, our algorithm ensures that the spatial location of objects is not ignored. The pair $A_1 - B_1$ gets more transactions (shown as black dots in Figure~\ref{fig_prox_remote}) than the second pair of objects. Therefore, the real situation is represented more accurately. Consider another example. Let spatial points $A_1$, $B_1$ and $C_1$ be pairwise neighbors (Figure~\ref{fig_intersect_a}). They are considered to form a clique by previous algorithms. However, as it can be seen in Figure~\ref{fig_intersect_b}, with certain buffer sizes it is possible that an actual intersection area of three buffers is relatively small. Furthermore, a scenario exists when there is no intersection of the three objects at all, although they form pairwise neighbourhood relationships, as it is illustrated in Figure~\ref{fig_intersect_c}. Our buffer-based framework is able to distinguish these cases. A varying number of transactions is derived from intersecting regions of multiple objects depending on distances between them and their buffer sizes.

\begin{figure*}[!t]
\centering
\begin{minipage}{1.3in}
\subfigure[$A_1$, $B_1$ and $C_1$ are pairwise neighbors]
{\includegraphics[width=1.3in]{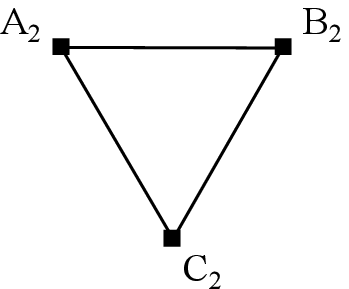}
\label{fig_intersect_a}}
\end{minipage}\hspace{0.2in}
\begin{minipage}{1.3in}
\subfigure[An intersection of $A_1$, $B_1$ and $C_1$]
{\includegraphics[width=1.3in]{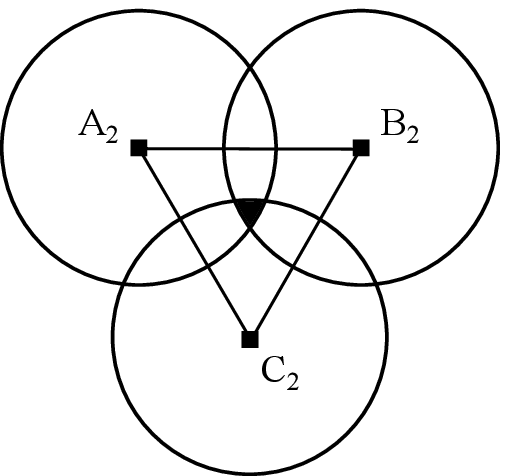}
\label{fig_intersect_b}}
\end{minipage}\hspace{0.2in}
\begin{minipage}{1.3in}
\subfigure[No intersection of $A_1$, $B_1$ and $C_1$]
{\includegraphics[width=1.3in]{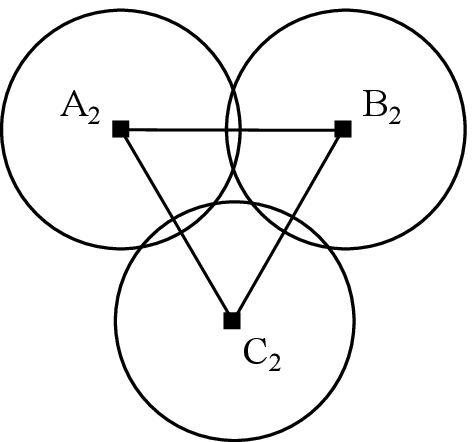}
\label{fig_intersect_c}}
\end{minipage}
\centering
\caption{Intersection of neighboring objects}
\label{fig_intersect}
\end{figure*}

\item Similar to classical frequent pattern mining applications where data can be certain (deterministic) or uncertain (probabilistic), spatial datasets can also exhibit uncertainty of feature existence in space. In other words, a probability of detecting an existence of a feature in a region closer to an observation point is higher than in regions situated farther from it. By taking into account uncertainty and including it in our framework, we believe that our model increases the accuracy of the results.
\end{itemize}

\section{Modeling Framework}
\label{mf}
A modeling framework that is used to handle and analyze data is an important part of any practical research. In theoretical studies it could be simplified in order to generalize a task and define algorithms that could be applied for a wide range of applications and domains. However, the usage of general approaches and algorithms may result in misleading or even wrong results. For example, a neighbourhood distance threshold is an important measure of an interaction and relationship between features. Obviously, a single distance threshold cannot accurately capture all possible relationships among features. In biology, various animal species have different home ranges, areas where they search for food. Rodents may require little space, while birds forage on wider regions. Another example can be derived from urban studies. Consider two points of interest, for example a shopping mall and a grocery store, situated in a distance exceeding some threshold. However if they are connected by a high quality road, they are more likely to be co-located than any other two points positioned seemingly close to each other but separated by some obstacles. To tackle this kind of application specific issues, most of the application domains, if not all, have their own nuances that must be taken into account, when performing any analysis or mining tasks, in order to get most accurate and significant results.

The motivating task of this paper, detecting co-locations of pollutant emission points and childhood cancer cases, has unique difficulties and challenges. A distribution of a pollutant in a region is not uniform and it could depend on several factors: types of pollutants, amounts of release, weather conditions (wind, precipitation), topography, etc. Various chemicals have different levels of harmfulness and toxicity. In addition, a pollutant concentration might be inversely proportional to the distance from the emitting point. These are only a few examples. We show how we tackled some of these problems such as pollutant amounts, wind speed and direction, and uncertainty of presence of chemicals. Certainly, we do not aim to reproduce complicated air pollution distribution models which require considering many other variables and parameters. Instead, our model gives a simple framework that attempts to simulate real scenarios while operating with available data.

The motivating application of this paper, detecting co-location patterns/rules of pollutant emission points and childhood cancer cases, has unique difficulties and challenges which need to be addressed. For example, a distribution of a pollutant, in a region surrounding the emission point of it, is not uniform and it could depend on several factors such as pollutant type, released amount, weather conditions (e.g. wind and precipitation), topography, etc. On the other hand various chemicals have different levels of harmfulness and toxicity. In addition, the concentration of some  pollutants might be inversely proportional to the distance from the emitting point. These are only a few examples of possible issues in our motivating application. Addressing these issues can be really important to the accuracy of our algorithm. For example, when defining buffers for spatial data points, an accurate model which can tackle the above mentioned issues is very important. In following we discuss how our modelling framework tackled some of these problems such as released amount, wind speed and direction, and the uncertainty in the existence of the chemicals. Certainly, we do not aim to reproduce complicated air pollution distribution models which require to consider many other variables and parameters. Instead, our model gives a simple framework that attempts to simulate real world conditions while operating with available data.
\subsection{Pollutant Amounts}
\begin{figure*}[!t]
\centering
\includegraphics[width=4.5in]{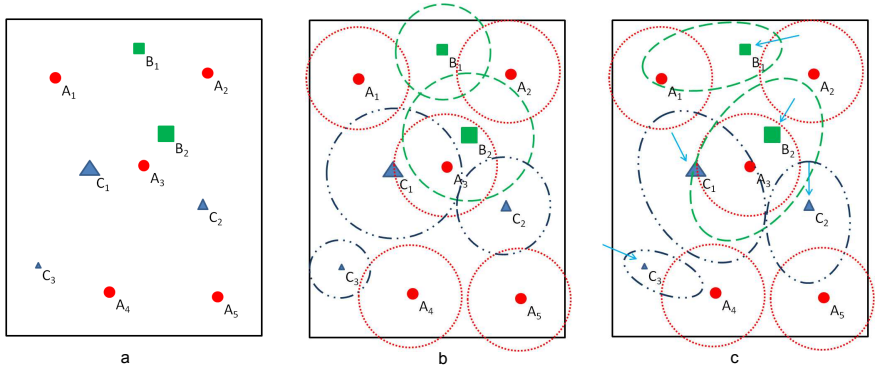}
\caption{Modeling framework usage examples: (a) an example spatial dataset (A - cancer, B and C - pollutants); (b) buffer sizes vary depending on the pollutant release amount; (c) buffer shapes change with the wind direction and speed (shown by arrows)}
\label{fig_framework}
\end{figure*}

The dataset on pollutants contains the data on estimated yearly releases of chemicals by industry, according to Canada's National Pollutant Release Inventory~\cite{npri}. For our research we use the average amount of released chemicals by a facility in a given year. The range of the average amount values varies from several kilograms to tens of thousands of tons; the minimum and maximum average yearly release in the dataset is 0.001 grams and 80,000 tons, respectively. Figure~\ref{fig_framework}~(a) displays an example dataset containing cancer points (feature $A$) and chemical points (features $B$ and $C$). On Figure~\ref{fig_framework}~(b) buffer zones around pollutant points are based on the amount of a release at that location. For example, instance $C_1$ has a larger zone affected by this source point than instance $C_3$ which has a smaller amount of emission. Buffer zones of cancer points are not changed.

For simplicity, we decided to take the maximal distance as the natural logarithm function of the release amount. This function gives a smooth curve which does not grow as fast as linear or root functions that give large numbers for heavier releases. Even though this technique oversimplifies the real world conditions of pollutant dispersion, it helps to make the results more precise. Other functions can be used to calculate the maximal distribution distance and they can depend on a type of a pollutant (a heavier chemical settles faster and on a shorter distance from a chimney) or a height of a chimney. An additional point that could be considered in future work is that the area very close to a chimney does not get polluted, and the higher is the chimney, the bigger is that region.

\subsection{Wind Speed and Direction}
The weather conditions and topographical features may affect the distribution of chemicals in the air. The examples of these factors are prevailing winds, precipitation, relative humidity, mountains, hills, etc. At the first step in this part of the modeling framework we include the wind speed and the prevailing wind direction on source points as variables of the model.

\begin{figure}[!t]
\centering
\includegraphics[width=4in,height = 1.1in]{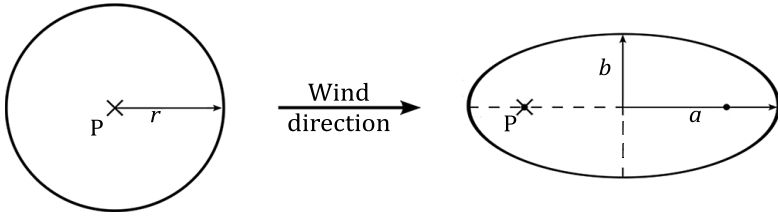}
\caption{A buffer circle around emission point $P$ is morphed into an ellipse}
\label{fig_circle_ellipse}
\end{figure}

Regarding the wind speed and direction, two situations are possible. First, a region where a facility is located is windless throughout the year. In this case, a pollutant is assumed to disperse in a circular region around the source point with a radius of a circle derived from a released amount as discussed in the previous subsection. However, a second situation is more frequent; there is nonzero wind speed with a prevailing wind direction. In this case we presume that the original distribution circle is morphed into a more ellipse-like region. Figure~\ref{fig_framework}(c) illustrates elliptical buffer regions; their forms are dependent on the wind speed and its frequent direction. Our calculations of the characteristics of an ellipse are based on the works by Getis and Jackson~\cite{Getis1971}, and Reggente and Lilienthal~\cite{Reggente2009}. The major axis of the ellipse is in the direction of the prevailing wind. We assume that the area polluted by a chemical when a wind is present is the same as when there is no wind. Therefore, the coverage area of the ellipse is kept equal to the area of the original circle. The source point can be placed on the major axis of the ellipse between the center and upwind focus; in our model we locate it in the middle of the segment between these two points. Figure~\ref{fig_circle_ellipse} shows an example of buffer transformation. The original circle buffer zone around the emission point $P$ is changed to an ellipse. As Figure~\ref{fig_circle_ellipse} shows that the center of the buffer (i.e. the source of the pollutant) becomes a foci of the ellipse.

Obviously, wind with a higher speed distributes chemicals to greater distances. Therefore, we need to include the wind speed value in the computations. The lengths of the major semi-axis $a$ and minor semi-axis $b$ are dependent on the wind speed and derived from the equations:
\begin{equation}a= r+\gamma |\vec v|,\end{equation}
\begin{equation}b= \frac{r^2}{a},\end{equation}
where $r$ is the radius of the original circle, $\vec v$ is the wind speed, and $\gamma$ is the stretching coefficient.

The larger a value of the stretching coefficient, the longer is the ellipse's major axis. In this work it is fixed at 0.3, but it could be changed and have a different value for each of pollutants. The calculation of the length of the semi-minor axis $b$ follows our assumption that the area of the ellipse is equal to the area of the original circle.

We improve our model by using elliptical buffer zones, which depend on the average wind speed and most frequent wind direction, instead of circular buffers. However, this is only a simplified model. Other factors which affect chemical distribution in air might be taken into account in future research to more accurately simulate real processes. As an example multiple alternating seasonal wind direction would affect the buffer shape to a more complex appearance.

\subsection{Wind Stations and Data Interpolation}
\begin{figure*}[!t]
\centering
\begin{minipage}{2.0in}
\subfigure[18 stations of Environment Canada]
{\includegraphics[width=2.0in,height = 2.6in]{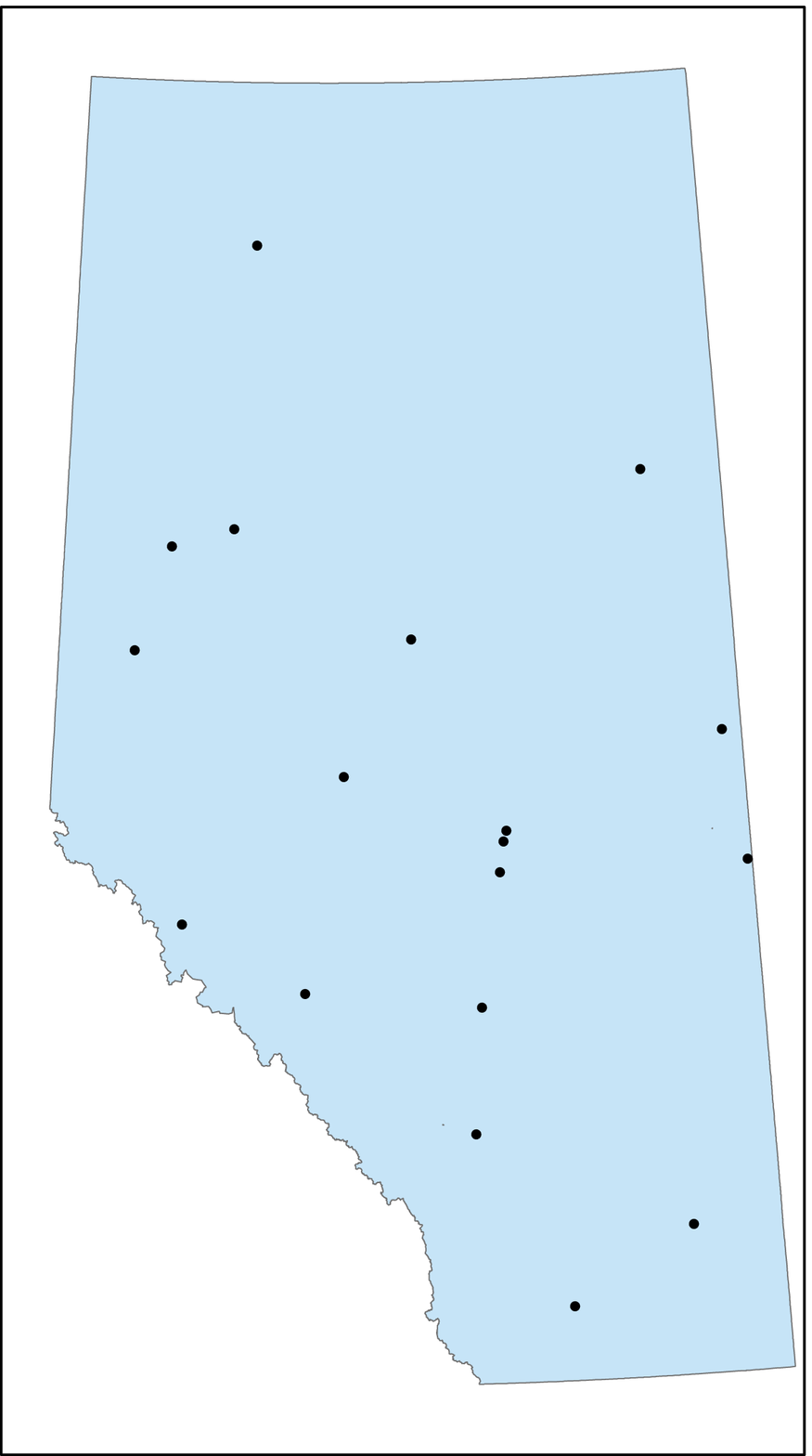}
\label{fig_wind_a}}
\end{minipage}\hspace{0.3in}
\begin{minipage}{2.0in}
\subfigure[156 stations of ACIS]
{\includegraphics[width=2.0in,height = 2.6in]{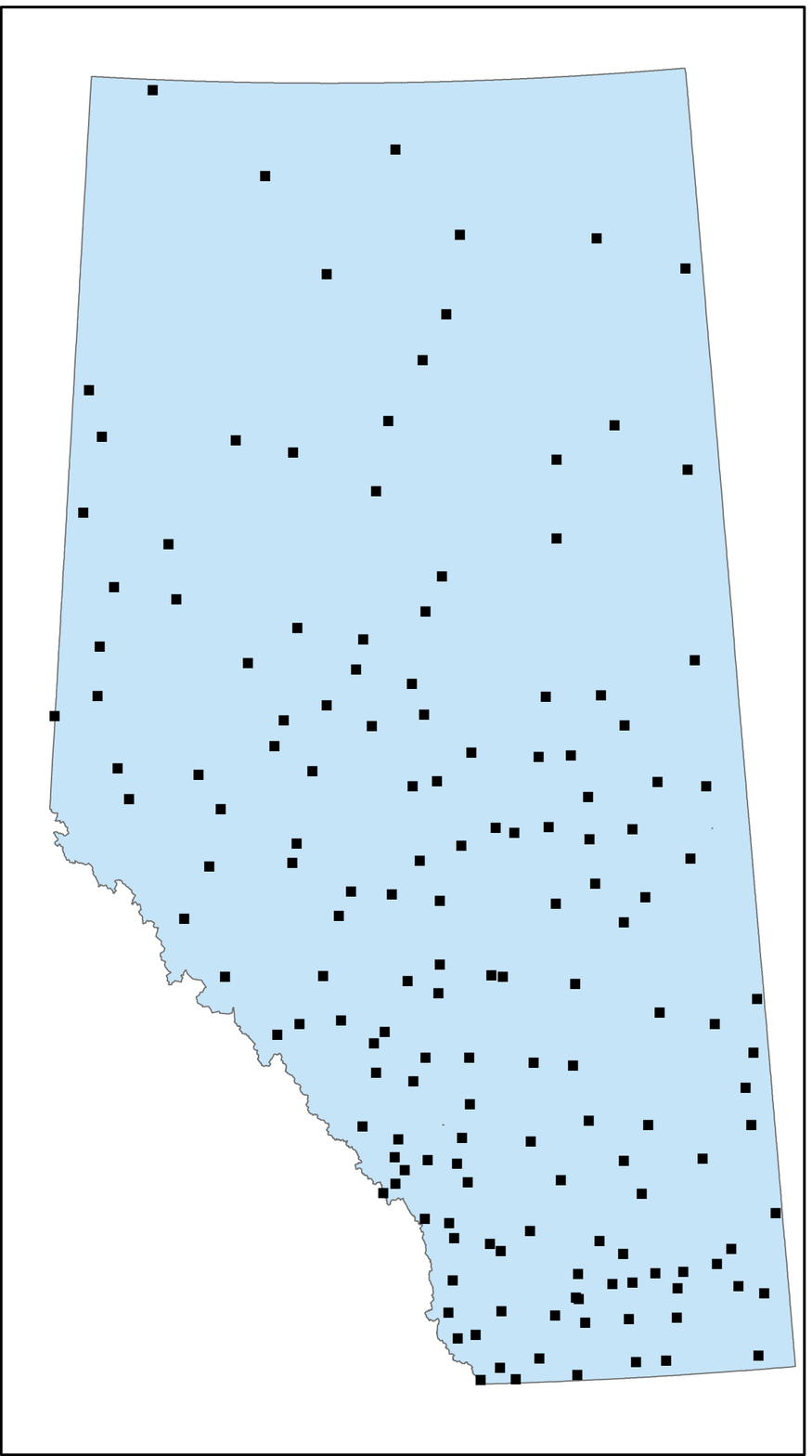}
\label{fig_wind_b}}
\end{minipage}
\centering
\caption{The monitoring stations in Alberta}
\label{fig_wind}
\end{figure*}

In order to get values of the wind speed and prevailing wind direction, an interpolation of wind fields between weather stations is used. The data of monitoring stations in Alberta comes from two sources. First, the data from 18 stations is obtained from Environment Canada~\cite{env_canada} which provides climate normals that are based on climate stations with at least 15 years of data between 1971 and 2000. The most frequent wind direction is a direction (out of possible eight directions) with the highest average occurrence count. Second, the data from 156 stations is derived from AgroClimatic Information Service (ACIS)~\cite{acis}. The locations of stations in Alberta are displayed in Figure~\ref{fig_wind}. The data provided by ACIS is daily from 2005 to 2011. In order to make the data consistent, the average wind speed and the most frequent wind direction are calculated using a method similar to the one used by Environment Canada~\cite{env_canada}. The average wind speed is simply the average value of this parameter for all available days. The wind direction is rounded to eight points of the compass. A direction with the highest count of daily observations is assigned as the most prevailing wind direction.

\begin{figure}[!t]
\centering
\includegraphics[width=3in, height = 3in]{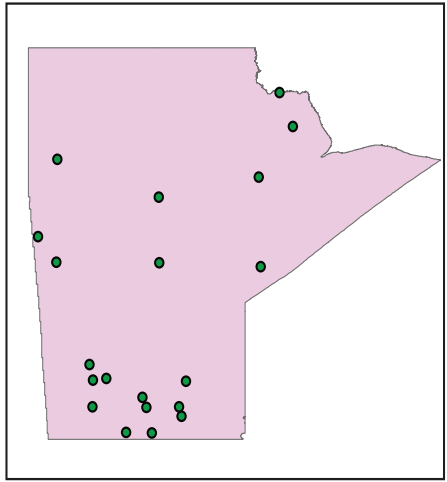}
\caption{The monitoring stations in Manitoba}
\label{fig_wind_manitoba}
\end{figure}

Unlike Alberta, the data of monitoring stations in Manitoba comes from only one source, 20 stations from Environment Canada~\cite{env_canada}, it also contains climate normals information from the years of 1971 to 2000. The locations of stations in Manitoba are displayed in Figure~\ref{fig_wind_manitoba}.

The climate normals from two sources in Alberta and one source in Manitoba are used to make interpolations in the ArcGIS tool~\cite{arcgis}. However, ArcGIS is restricted to linear surface interpolations and the wind direction is a nonlinear attribute. In linear systems (e.g., the number of sunny days or days with precipitation) there is only one unique path when moving from one number to another, for example, if we want to move the temperature from 37$^{\circ}$C to 40$^{\circ}$C, the only path is to linearly increase the first degree. On the other hand, nonlinear systems may have several paths. For example, there are clockwise and counter-clockwise directions to move from 90$^{\circ}$ to 270$^{\circ}$: through 0$^{\circ}$ or 180$^{\circ}$. These directions go from one point to the second but both are unique. Therefore, linear interpolations lead to wrong results when deployed directly to non-linear systems.

Interpolation of wind fields requires a technique that considers the non-linear nature of the wind direction attribute. A transformation is done according to the work by Williams~\cite{Williams1999}. The wind speed and wind direction from each monitoring station is represented as a vector with the magnitude $S$ (wind speed) and direction $\theta$ (wind direction). The vector is divided into axial components $X$ (northern wind) and $Y$ (eastern wind):
\begin{equation}X=S \sin \theta,\end{equation}
\begin{equation}Y=S \cos \theta.\end{equation}

Based on these two components, two ArcGIS surface interpolations are created. The type of interpolation used is spline. As a result we get two grids: for northern $X'$ and eastern wind $Y'$. The magnitude of the vector, the wind speed $S'$, is computed as:
\begin{equation}S'=\sqrt{X'^2+Y'^2}.\end{equation}

\begin{figure}[!t]
\centering
\includegraphics[width=2.5in, height = 2.5in]{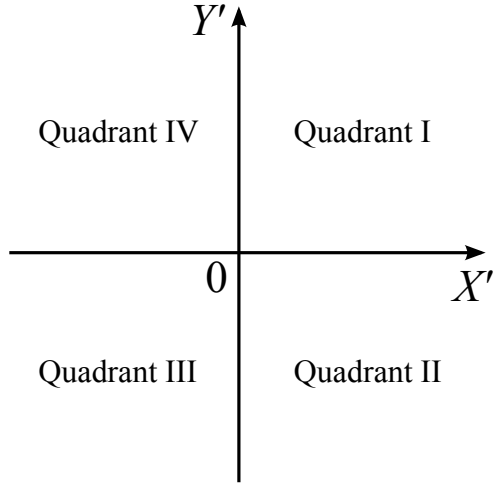}
\caption{Four quadrants defined by the signs of values of $X'$ (northern wind) and $Y'$ (eastern wind)}
\label{fig_quadrants}
\end{figure}

The calculation of wind direction angle $\theta'$ is more complicated. From geometry, the wind direction is calculated as $\theta' = \tan^{-1}(Y'/X')$. However, the inverse tangent is defined only for values between -90$^{\circ}$ and 90$^{\circ}$ and it is only half of our domain. Therefore, each of the four quadrants of our domain (the quadrants are shown in Figure~\ref{fig_quadrants}) requires its own formula~\cite{Williams1999}:
\begin{equation}Quad~I:\theta' = \tan^{-1}(X'/Y'),\end{equation}
\begin{equation}Quad~II:\theta' = \tan^{-1}(Y'/X')+90^{\circ},\end{equation}
\begin{equation}Quad~III:\theta' = \tan^{-1}(X'/Y')+180^{\circ},\end{equation}
\begin{equation}Quad~IV:\theta' = \tan^{-1}(Y'/X')+270^{\circ}\end{equation}

As a result we get interpolated values of wind speed and wind direction for each point of the studied space.

\subsection{Data Uncertainty}
 The dispersion of a pollutant in a distribution region is not uniform, and intuitively the concentration of the pollutant near a chimney is higher than at a border of the dispersion region. Furthermore, pollutants are subject to decay and deposition processes. In other words, it is more likely that people living near an emitting facility are exposed to higher levels of pollutants than people who live kilometers away from the facility. Therefore, presence of a chemical in a given point is uncertain and a probability of detecting it depends on a distance from the point to the emission source. This dependency is inversely proportional. For example, in Figure~\ref{fig_grid_c} the probability of detecting $A$ at the point $gp_1$ is lower than at the point $gp_2$.

\begin{figure*}[!t]
\centering
\begin{minipage}{2.0in}
\subfigure[Categorical function.]
{\includegraphics[width=2.0in]{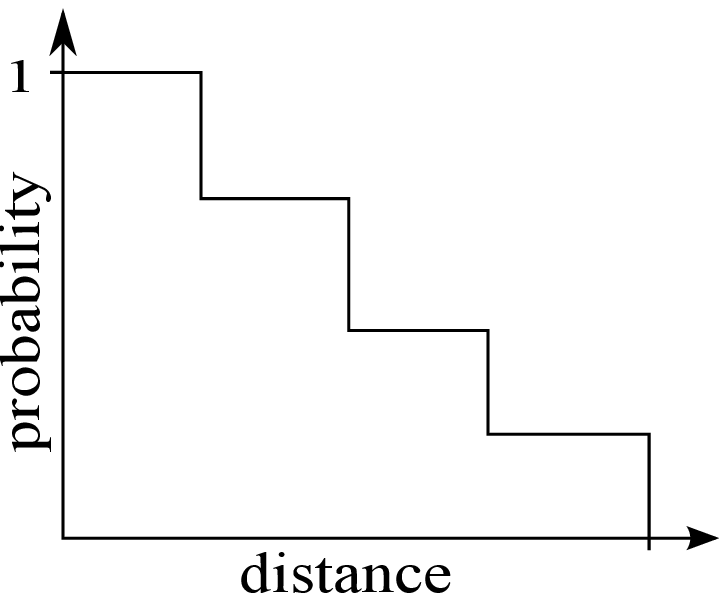}
\label{fig_uncertain_func_a}}
\end{minipage}\hspace{0.3in}
\begin{minipage}{2.0in}
\subfigure[Linear function.]
{\includegraphics[width=2.0in]{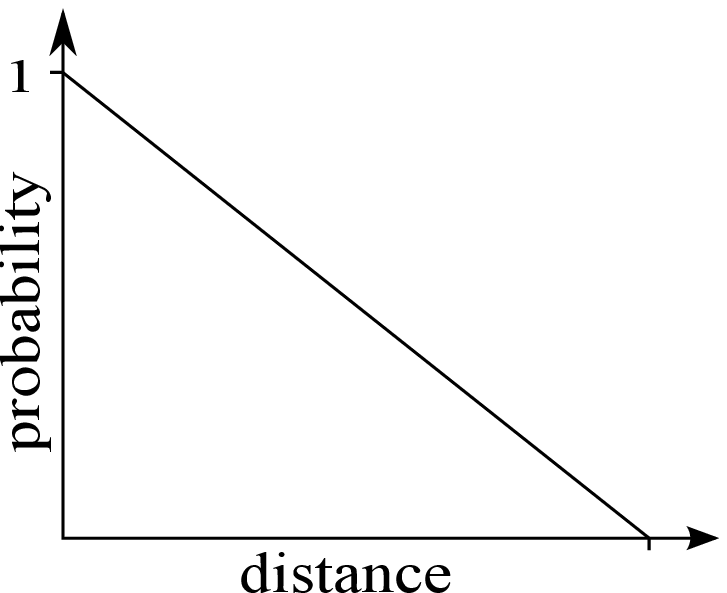}
\label{fig_uncertain_func_b}}
\end{minipage}\hspace{0.3in}
\begin{minipage}{2.0in}
\subfigure[Curve function ($\frac{\cos{\pi x}}{2}+0.5$).]
{\includegraphics[width=2.0in]{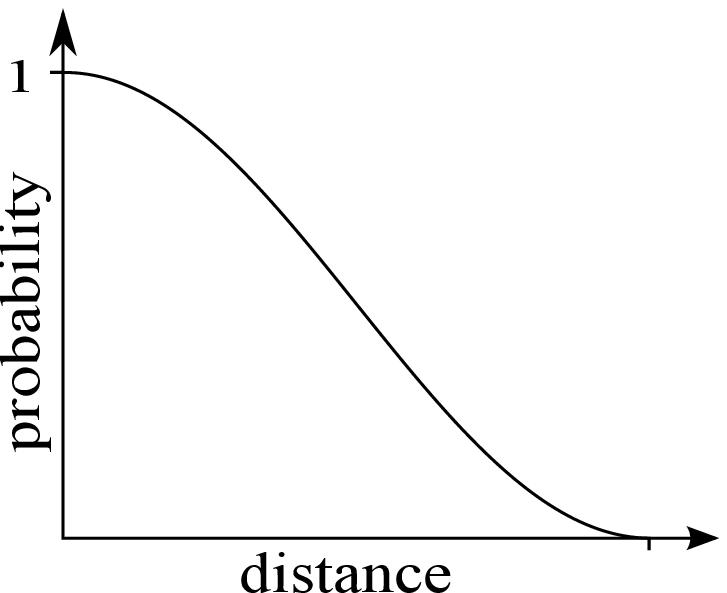}
\label{fig_uncertain_func_c}}
\end{minipage}
\centering
\caption{Examples of functions that can be used to represent the dependency of the pollutant presence probability on the distance to the source point}
\label{fig_uncertain_func}
\end{figure*}

Various functions can be used to determine the dependency of the pollutant presence probability in a given point on the distance to the emitting facility. For instance, with a categorical function (Figure~\ref{fig_uncertain_func_a}), we assign probabilities according to distance ranges, e.g., 1.0 for 0-2 km from the facility, 0.75 for 2-4 km, 0.50 for 4-6 km, etc. Another example is a linear function (Figure~\ref{fig_uncertain_func_b}) which can be represented as $1 - x'/x$, where $x'$ is the distance from a given point to the facility and $x$ is the maximal distance where pollutant distributes. In this work we use a third example, the curve function (Figure~\ref{fig_uncertain_func_c}), which is derived from the cosine function, $p = \frac{\cos{\pi x}}{2}+0.5$. With this function the probability decreases slowly with the increasing distance. Then, it starts declining more linearly, and at the end slows down again. We believe that the curve function models the real-life pollutant behavior more accurately than the other two methods.

These three examples are only some of possible curves that can be used to model pollutant distribution within buffer zones. However, other functions could be used in order to improve the accuracy of the results. They could depend on the types of chemicals. For example, denser chemicals may settle out in a region closer to the emitting facility, while only small amounts would reach places at medium and far distances.

\begin{figure}[!t]
\centering
\includegraphics[width=4.5in]{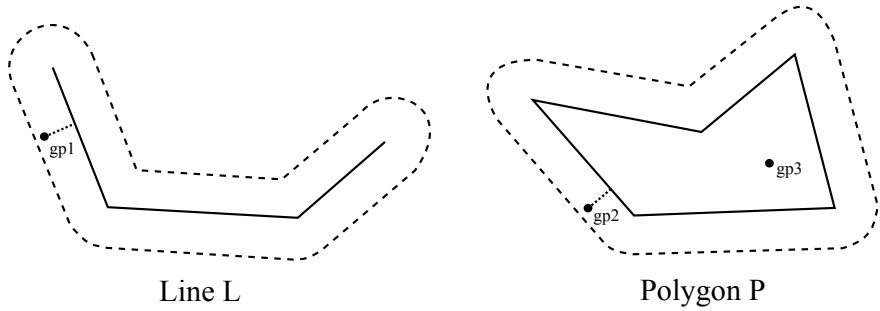}
\caption{Defining a distance to an object in datasets with polygons and lines}
\label{fig_uncertainty}
\end{figure}

In the case of datasets which contain other types of spatial objects (i.e. lines and polygons), grid points intersecting a line or located inside a polygon are assigned a feature presence probability of one for the corresponding feature. For an example see point $gp_3$ in Figure~\ref{fig_uncertainty}. On the other hand, uncertainty for grid points positioned in buffer zones depends on the shortest distance from the point to the line or polygon. Points $gp_1$ and $gp_2$ in Figure~\ref{fig_uncertainty} are located in buffer zones of line $L$ and polygon $P$ respectively. Existential probabilities at these points are computed using shortest distances to respective spatial objects.

\section{Experimental Evaluation}
We evaluate our method on both real and synthetic datasets to measure the performance and the effectiveness of the approach. In the following discussion we explain our datasets, results and analyze the results on various aspects such as grid granularity, computational complexity, etc. We also compare our approach with a baseline approach to understand how effective our approach is in identifying not only highly prevalent rules, but also statistically significant ones.
\subsection{Real Data}
We conduct experiments on two real datasets which contain data on pollutant emission points and childhood cancer cases in the provinces of Alberta and Manitoba, Canada. The sources of the databases are the National Pollutant Release Inventory (NPRI, the data is publicly available)~\cite{npri}, and the provincial cancer registries. The National Pollutant Release Inventory (NPRI), established in 1992, is the national Pollutant Release and Transfer Register (PTRT) of Canada. The information on pollutants is taken for the period between 2002 and 2007 and contains the type of a chemical, location of release, and average amount of release per year. In order to get reliable results, the chemicals that had been emitted from less than three facilities are excluded from the dataset. There are 47 different chemicals and 1,442 pollutant emission points in Alberta; 26 different chemicals and 545 pollutant emission points in Manitoba, several chemicals might be released from the same location. The number of cancer points (the centroids of postal code regions where children lived when cancer was first diagnosed) are 1254 and 520 in Alberta and Manitoba, respectively.

Environmental pollutants are suspected to be one of the causes of cancer in children. However, there are other factors that could lead to this disease (genetic susceptibility, parental exposure to chemicals or radiation, parental medical conditions, etc.). Considering this fact, we attempt to find ``correlations" rather than ``causalities". The results are currently under careful evaluation by domain experts in our interdisciplinary team. It suffices to mention, however, that some surprising rules were discovered indicating significant association between groups of chemicals, not categorized individually as carcinogens, and children cancers. Additionally, we identified rules co-locating cancer cases with a pair of chemicals, one known as carcinogen and another with no-carcinogenic properties. Since the carcinogen alone did not correlate with cancer, the occurrence of those pairs of chemicals suggests a ``catalyzing" effect by the non-carcinogen.

We are interested in co-location rules of the form $Pol \to Cancer$, where $Pol$ is a set of pollutant features and $Cancer$ is a cancer feature. The expected confidence is used as a prevalence measure. The distance between points in a grid is 1 km; the change in the grid granularity is also evaluated. The number of simulations (randomized datasets) for the statistical test is set to 99, so that with the observed data the denominator in Equation~(\ref{eq_pi_value}) is 100. The level of significance $\alpha$ is set to 0.05. The size of an antecedent of candidate rules is up to three. Larger candidates have low support values due to the fact that the average number of features in a transaction in the experiment is up to 3.

The randomized datasets that are used in the statistical test are generated as follows. Pollutant emitting facilities are not random and usually located close to regions with high population density, while they are not present in other places (e.g., in protected areas). Based on this observation, we do not randomize pollutant points all over the region, but instead keep locations of facilities and randomize pollutants within these positions. Among 1,254 cancer points in Alberta, 1,134 were located within ``urban" municipalities (cities, towns, villages, etc.) and the rest were diagnosed in ``rural" areas. For the cancer cases in Manitoba, 400 out of 520 were located in urban areas, while the rest, 120 in rural areas. In order to have the randomized cancer occurrence rate close to the real-world rate, we keep the number of cancer feature instances positioned in ``urban" /``rural" regions the same as in the real dataset. The number of random cancer cases placed within each ``urban" municipality is directly proportional to the number of children counted in the 2006 census~\cite{census}. The rest of the cancer cases in the real datasets, are randomly placed in rural regions on the map of Alberta and Manitoba. The detailed information of pollutants and cancers in Alberta and Manitoba is displayed in Table~\ref{table_information}.

\begin{table}[!t]
\caption{Detailed pollutants and cancers information in Alberta and Manitoba}
\label{table_information}
\centering
\begin{tabular}{|c|c|c|c|c|}\hline
Dataset  & $\#$Pollutants & $\#$Cancers & $\#$Cancers (urban) & $\#$Cancers (rural)\\\hline
Alberta  & 1,455 & 1,254 & 1,120 & 134 \\ \hline
Manitoba & 545   & 520   & 400   & 120 \\ \hline
\end{tabular}
\end{table}
\subsubsection{Comparison with Certain Data Method}
We compare the results of our uncertain data method (UM) with the results of a method using certain deterministic data (CM) where existential probabilities are not stored as a part of a transactional database. As an interestingness measure in the CM we use the confidence $Conf(Pol \to Cancer)$, which is a fraction of transactions containing all features in $Pol$ that also include the cancer feature.
\begin{equation}Conf(Pol \to Cancer)=\frac{Sup(Pol \cup Cancer)}{Sup(Pol)}.\end{equation}

In Alberta, the number of significant co-location rules detected by both UM or CM together is 496; from these 204 rules are found by both methods, 278 rules are identified only by UM, and 14 rules by CM. In Manitoba, the number of detected significant co-location rules by UM and CM is 362 and 263, respectively, and these two methods have 232 common rules. From these two datasets, we can find that the UM method covers most of the significant co-location rules detected by the CM method. Moreover, UM captures more statistically significant rules that can not be mined by CM otherwise. The difference in the results could be explained by the fact that our approach deals with probabilities of feature presence in transactions rather than with deterministic values. It considers not only the presence of a feature in a transaction but also distances from grid points to pollutant features and cancer cases. The grid points that are situated closer to spatial instances are given more weight than points located relatively farther.

Some of the co-location rules discovered by the uncertain method have a low level of $ExpSup(Pol)$ or $ExpSup(Pol \cup Cancer)$. For example, 348 out of 482 rules in the Alberta dataset and 193 out of 362 rules in the Manitoba have $ExpSup(Pol \cup Cancer)$ less than 1. It means either a low number of transactions or a relatively long distance from grid points. Although with a low $p$-value ($\le 0.05$) they have an expected confidence level higher than in most randomized datasets, domain experts might not be interested in these co-location rules. In that case, a threshold on the expected support might be introduced to the model for detection of significant co-location rules. This threshold should not be set too high, so that the algorithm does not miss some of the interesting co-location rules or patterns with rare features.
\subsubsection{Comparison with Support/Confidence based Method}
Comparison of our approach with other methods existing in the literature is not straightforward. Especially because most of the existing methods expect boolean spatial features, are not capable of handling uncertainty, require to define reference features, require to define neighborhood relationships based on a single threshold value, etc. Our proposed approach attempts to address these limitations and the way we model the spatial data and features to represent the real world conditions are hard to be captured by these existing models. Hence we implement a na{\"i}ve co-location mining method based on the traditional support/confidence framework but is capable of gaining the advantage of our underlying data modelling framework. More specifically this approach takes advantage of the grid-based transactionization and uncertainty information of features.

To compare the effectiveness and the robustness of our approach the implemented na{\"i}ve approach acts as a baseline method which can obtain a set of significant rules based on the traditional confidence/support framework. This na{\"i}ve method uses our Algorithm~\ref{alg_transaction} to compute the transactions, then compute the prevalence measure for each candidate pattern as in Algorithm~\ref{alg_main} (see steps 4-9). In the present case we are interested in mining association rules, hence we used $ExpConf(X \to Y)$ as the prevalence measure. Once we computed the expected confidence for all the candidate patterns, we calculated the average expected confidence for all the candidate rules. We used this as a cut-off threshold to prune the candidate rules. In the case of Certain Method this pruning step was able to prune 12239 rules from a candidate rules set size of 17343, obtaining a rules set size of 5104 as important rules. In this case the expected confidence threshold was $62\%$. In the case of Uncertain Method the pruning step was able to prune 12326 rules from a candidate rules set size of 17343, obtaining a rules set size of 5017 as important rules. In this case the expected confidence threshold was $46\%$.

We tested the statistical significance of these rules using the statistical tests we explained in Section~\ref{ad} (see steps 10-22 in Algorithm~\ref{alg_main}) and obtained 467 rules are statistically significant from the rules obtained from both Certain and Uncertain methods.

In the previous discussion we explained that CM and UM together identified 496 rules as statistically significant. We observed that all the 467 rules detected under the expected confidence based na{\"i}ve approach we implemented above are also obtained by our approach. In addition to that our approach was able to identify 29 more rules as statistically significant. This result emphasizes the key advantage of our approach: our approach is not only capable of detecting highly prevalent rules, but is also capable of detecting rules which may not be highly prevalent but statistically significant.
\subsubsection{Effect of Randomization}
As mentioned above, in the randomized datasets, we take an intuitive reasonable strategy by randomizing the pollutant emitting facilities and cancer cases. We randomize the pollutant emitting facilities in the areas with high density population, and randomize the cancer cases according to the population distribution in ``urban" and ``rural" areas. Moreover, the number of cancer cases in ``urban"/``rural" regions are the same as the real dataset. We compare this randomized strategy with other two randomized methods: randomizing the pollutants only and randomizing cancer cases only.

First in the Alberta dataset, it is discussed above that when we randomize both pollutants and cancers, we get 482 significant co-location rules by UM. Then we try to fix the cancer cases as the real dataset and only randomize the pollutant emitting facilities among the high density population regions. Up to 710 rules are reported under this setting. However, when we fix the pollutant facilities and only randomize the cancer cancers, only 127 rules are detected.

Then in the Manitoba dataset, we found 362 significant rules with both pollutant facilities and cancer cases randomized. 243 and 314 significant rules are reported when we only randomize the pollutant facilities and only randomize cancer cases, respectively.

All of three randomize strategies ensure that in the randomized datasets, the pollutant emitting facilities and the cancer cases are independent from each other. Knowing which randomization strategy is the best is still under investigation. All the patterns and rules discovered with the different strategies are being manually evaluated by domain experts. This detailed assessment would provide insights on which strategy would be best. However, in our current experiments, we take the first randomization strategy, by randomizing both the pollutant facilities and cancer cases which is more reasonable. This seems reasonable because it fairly randomizes all types of spatial data points. In the future, based on the suggestions from our domain experts, we might choose a different randomization strategy.

\subsubsection{Effect of the Grid Granularity}
As already mentioned, a granularity of the grid (a distance between grid points which affects the number of points per unit of space) is crucial for the result accuracy. A great distance between grid points may lead to the omission of some regions of the space especially when the average buffer distance is short. On the other hand, when the distance between points is too small, more transactions are derived by the algorithm. Decreasing the distance by a factor of two increases the transaction set size approximately by four times. Therefore, more computation needs to be done during the statistical test step. The grid resolution might be set up depending on the average buffer size.

For the Alberta dataset, in addition to the grid with a distance of 1 km between its points, we conduct two experiments with 2 and 0.5 km grids. As mentioned above, the algorithm reports 482 significant co-location rules with 1 km grid. With 2 km granularity 547 rules are detected from which 335 are present in both 1 and 2 km result sets, and 212 are unique for 2 km grid. The difference means that 2 km distance between grid points is too long for our dataset, where the average buffer size is 7.3 km, and its accuracy is comparatively low due to the smaller number of transactions which is not sufficient to capture intersections of instance buffers accurately. The 0.5 km granularity grid reports 472 co-location rules as significant. From these, 426 are found with both 1 and 0.5 km grids, and 46 rules are identified only by 0.5 km grid. As we can see, the difference between 0.5 km and 1 km result sets is smaller than that between 1 km and 2 km grids. As the distance between points in a grid decreases, the accuracy of the results improves.

We also conduct three sets of experiments with different grid granularity, 0.5 km, 1 km and 2 km on the dataset of Manitoba. With 1 km grid granularity, we detect 362 rules significant co-location rules. When the grid distance is increased to 2 km distance, 280 rules are reported and a large portion of them (271) rules also appear in 1 km result set. 364 significant co-location rules are found with 0.5 km grid granularity. Among these 364 rules, 356 rules are found in both 1 km and 0.5 km grid granularity. As can be observed, the difference between 1 km and 0.5 km grid distance is very small, and 1 km grid distance can cover most of significant rules detected by both 2km and 0.5km grid distance. Therefore, in our experiment, we choose 1km grid distance without loss of accuracy and efficiency. The industry emitted chemicals in Alberta and Manitoba are not the same, indeed they overlap very lightly. Therefore, we did not compare the set of significant co-location rules discovered in both datasets.

A closer analysis on the rules obtained by the CM method of our algorithm with Alberta dataset under the 0.5, 1 and 2km gird granularity measures provides some interesting insights on choosing a better granularity measure. We conduct this analysis focussing our concerns on two different aspects of the performance: accuracy, in terms of the p-value and efficiency, in terms of the execution time. Table~\ref{gridgrantime} shows when the grid granularity is decreased, how the number of transactions increases along with the execution time. This clearly shows when the grid granularity is small the number of transactions is huge leading to a larger execution time.

\begin{table}[t!]
\centering
\caption{Execution time, number of transactions and number of rules with respect to different grid granularity measures}
\label{gridgrantime}
\begin{tabular}{|l|l|l|l|l|}
\hline
Grid Granularity (km) & \begin{tabular}[c]{@{}l@{}}Number of\\ Transactions\end{tabular} & \begin{tabular}[c]{@{}l@{}}Execution\\ Time\end{tabular} & Rules \\ \hline
0.5              & 443653                                                           & 286 min 27 s                                             &   199           \\ \hline
1                & 129151                                                           & 110 min 33 s                                             &   218           \\ \hline
2                & 32255                                                            &               57 min 7s                                           &   273           \\ \hline
\end{tabular}
\end{table}

\begin{figure}[!t]
\centering
\includegraphics[scale = 0.6]{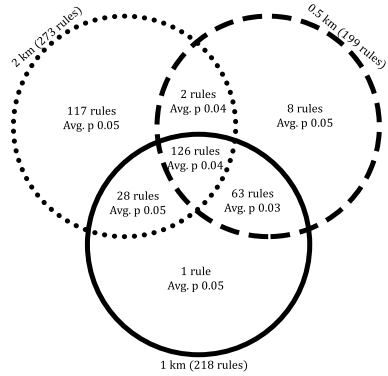}
\caption{Number of rules detected under various grid granularity measures (i.e. 0.5, 1 and 2 km) with the CM  method in the Alberta dataset and the average p-value of those rules}
\label{fig_venn}
\end{figure}

On the other hand, as depicted in Figure~\ref{fig_venn}, rules detected with the grid granularity of 1km shares all of its rules except one with the rules sets detected under 2 and 0.5km grid granularity. Out of these the rules shared with the rules set from 0.5km gird granularity has the lowest p-value (0.03) suggesting that it least complies with the null hypothesis. This indicates that the rules set commonly detected  under both 1km and 0.5km grid granularity are better than any other rule set. Figure~\ref{fig_venn} explains that the rule set detected under 1km grid granularity was able to capture most of the rules with better p-value from both the other rules set detected under 2 and 0.5 km grid granularity. This emphasizes that 1km could be a better choice when it comes to choosing a grid granularity because it captures most of the rules with better p-values while maintaining an efficient program execution. It can also be noticed that the all three rules set from different gird granularity measures shares 126 rules with each other. This common rule set is consistent irrespective of the grid granularity and shows better p-value, hence might be valuable for further studies.

\subsubsection{Effect of The Filtering Technique}
\begin{figure*}[!t]
\centering
\begin{minipage}{2.4in}
\subfigure[Filtering in Alberta dataset]
{\includegraphics[width=2.4in]{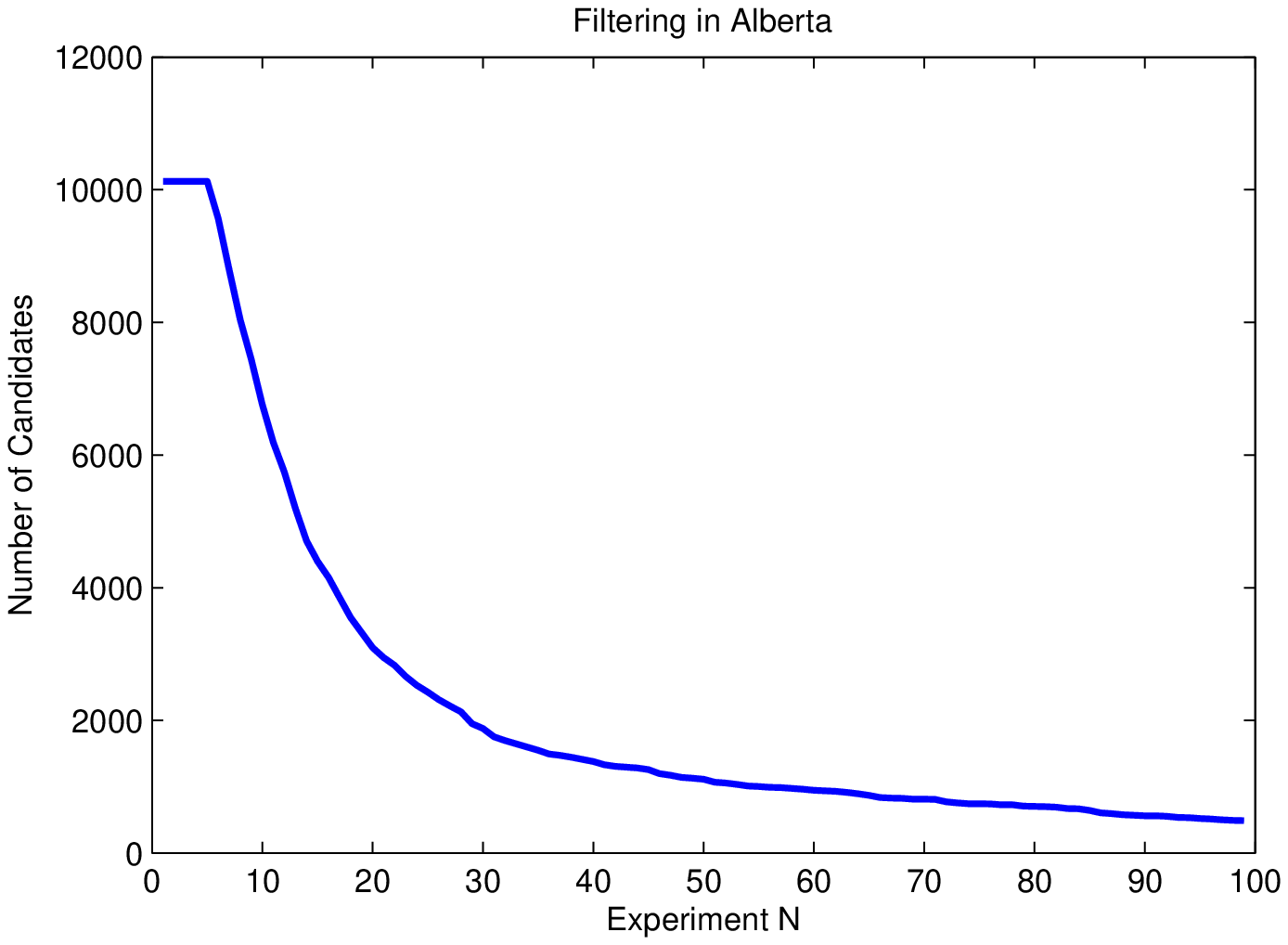}
\label{alberta_filtering}}
\end{minipage}\hspace{-0.2in}
\begin{minipage}{2.4in}
\subfigure[Filtering in Manitoba dataset]
{\includegraphics[width=2.4in]{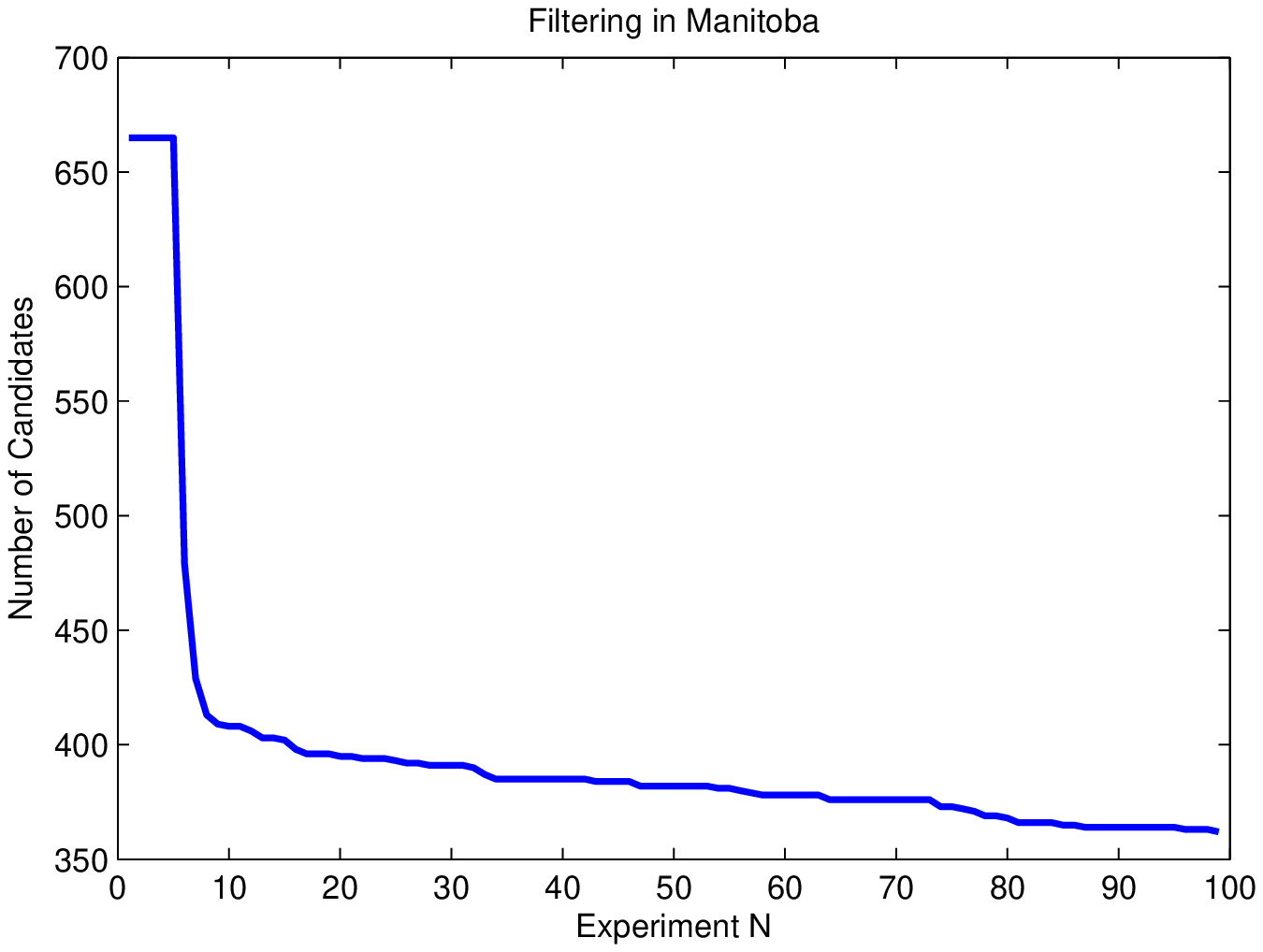}
\label{manitoba_filtering}}
\end{minipage}
\centering
\caption{The number of candidate rules evaluated in each simulation run with filtering technique}
\label{filtering}
\end{figure*}

In the Alberta dataset, the number of candidate co-location rules in the experiment (i.e. not yet determined as significant) is 17,343 (co-locations with the antecedent size up to three). With the na\"{\i}ve approach all candidates are checked in each of the 99 simulation runs which results in a large amount of computation. After the exclusion of rules with zero-level confidence, 10,125 candidates remain in these two datasets which also form a big set. Figure~\ref{alberta_filtering} show that the usage of the second filtering method (the exclusion of candidates which $p$-value) passed $\alpha$ during the evaluation of randomized datasets) considerably reduces the amount of computation. In the first simulation run the confidence value is computed for 10,125 rules, while 3,098 candidate rules are checked in the 20th simulation, and only 482 candidates are evaluated in the last run.

While in the Manitoba dataset, the total number of the candidate co-location rules with up to three pollutants is 2,951. Only 665 rules are left if we exclude the zero-level confidence co-location rules. It means that the na\"{\i}ve method can already help us prune a large amount of unnecessary rules. Figure~\ref{manitoba_filtering} shows the usage of the second filtering method. In the first simulation run the 665 rules are checked, and the candidate rules reduce to 408 in the 10th simulation which reduce around $\frac{1}{3}$ candidate rules, and in the last simulation only 362 candidates are checked.

\subsubsection{Effect of the Number of Simulation Runs on the Run Time}
\begin{figure*}[!t]
\centering
\begin{minipage}{2.4in}
\label{fig:compcomp}
\subfigure[Certain Method]
{\includegraphics[width=2.4in]{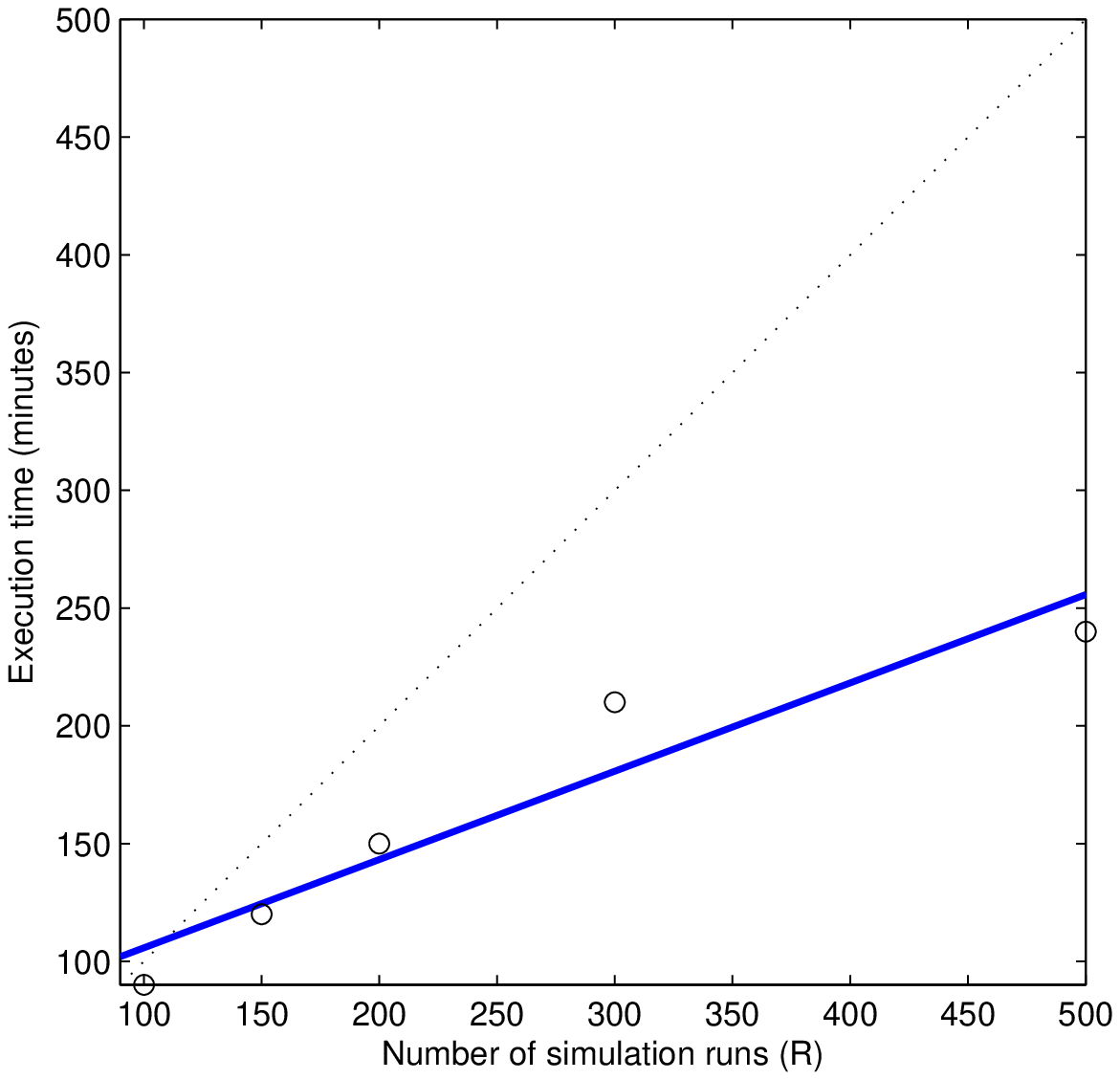}
\label{fig:compcomp1}}
\end{minipage}\hspace{-0.2in}
\begin{minipage}{2.4in}
\subfigure[Uncertain Method]
{\includegraphics[width=2.4in]{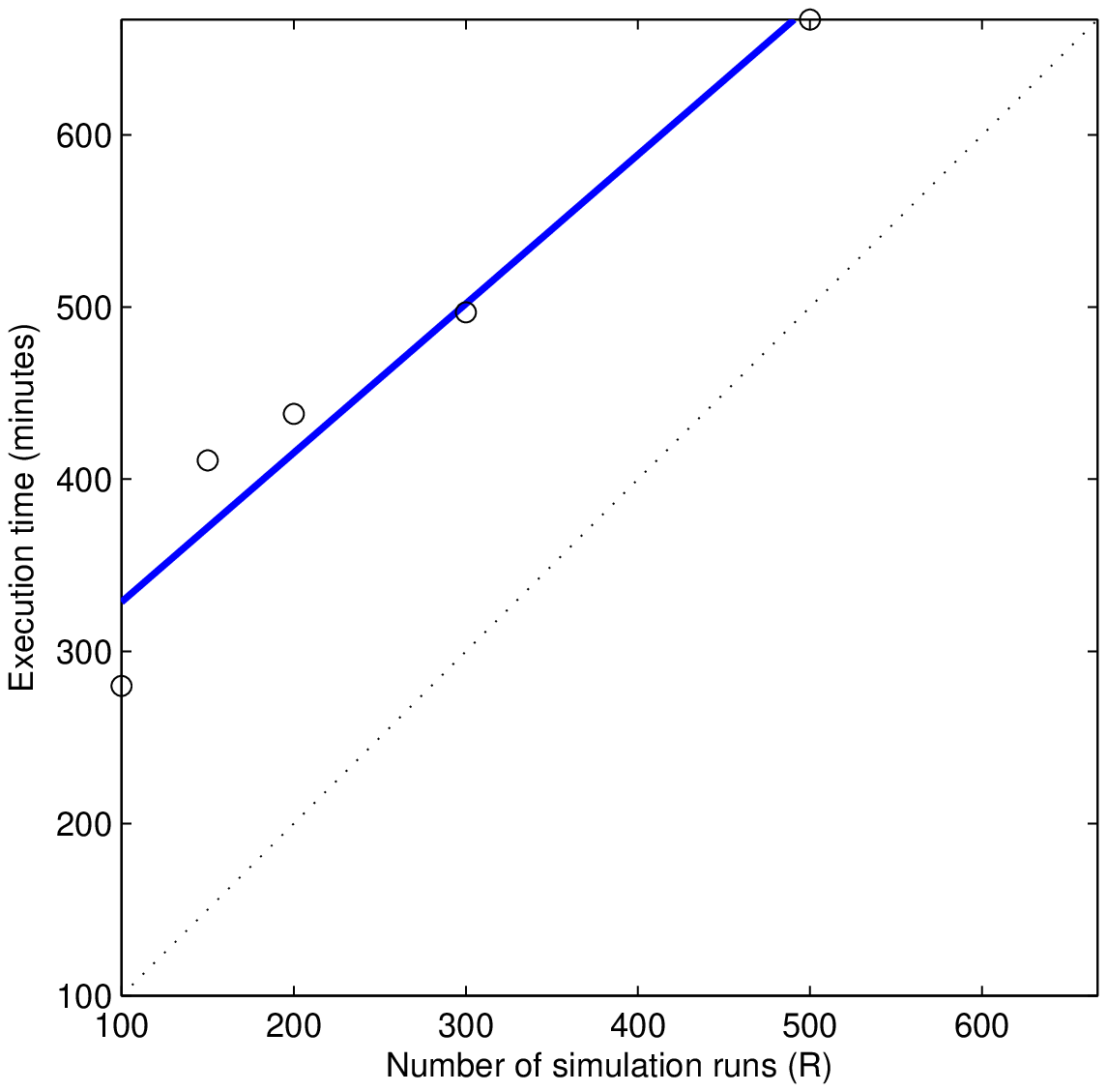}
\label{fig:compcomp2}}
\end{minipage}
\centering
\caption{Plots illustrating the execution time relative to the number of simulation runs}
\label{fig:exectime}
\end{figure*}

As explained in Section~\ref{cc}, Algorithm~\ref{alg_main} has a run time complexity of $O(n+m)$+O($k_3|f|^d$)+O(R$k_3|f|^d$), where \textit{n} is the number of chemical emission points, \textit{m} is the number of cancer cases, R is the number of simulation runs, $k_3$ is the number of grid points, $|f|$ is the size of the feature set $f$ and $d$ is the size of the largest item set. According to this if all the other variables except R are fixed and considered as constants the algorithm should have a linear time complexity with respect to the number of simulation runs (i.e. R). Figure~\ref{fig:compcomp1} and Figure~\ref{fig:compcomp2} represent the run time collected for both certain and uncertain methods on the province Alberta's datasets. A simple regression analysis reveals that the best line fits for the run time of the CM version of Algorithm~\ref{alg_main} is $y=3/8x+105/2$, as depicted in Figure~\ref{fig:compcomp1}. A similar regression analysis reveals that the best line fits for the run time of the UM version of Algorithm~\ref{alg_main} is $0.87x+240$, as depicted in Figure~\ref{fig:compcomp2}. This verifies the results from the previous analysis in Section~\ref{cc}, that run time has a linear complexity with respect to the number of simulation runs.

\subsection{Synthetic Data}
We conduct experiments on synthetic datasets to demonstrate that our framework can discover a correct set of co-location rules. In addition, we show that our transaction-based method takes into account spatial context and information.

\subsubsection{Discovery of Co-Location Rules}

In order to evaluate our algorithm on the synthetic data we generate a dataset and attempt to emulate the real-world information. Similar to the real dataset, it contains point features that appear in the antecedent part of co-location rules (features $C$), and a disease feature $D$ as a consequent. The study region is a 100$\times$100 unit square. The buffer size is 1 unit. We simulate 7 chemicals by using $C=\{C_{1},...C_{7}\}$, and $D$ is simulated as the cancer. The features $C_1$ and $C_2$ have 20 instances each and they are associated with each other. The features $C_3$ and $C_4$ have 30 points each; 20 of them are associated with each other, while remaining 10 instances are placed randomly. These two pairs represent co-located chemicals. The disease feature $D$ is positively associated with sets $C_1 \cup C_2$, $C_3 \cup C_4$, and with 30 out of 40 instances of feature $C_5$. It has no association with the feature $C_6$ (30 instances), and negatively correlated with $C_7$ (30 points), so that no pair of instances $D$ and $C_7$ are neighbors. In addition there are 30 disease cases spread randomly. We look for co-location rules of the form $C \to D$. In 99 randomized datasets all eight features are distributed randomly with no association (neither positive nor negative) with each other.

The significant co-location rules with $p$-value $\le 0.05$ are shown in Table~\ref{table_synt}. As expected, rules 5, 6, and 11 are reported as significant because they have strong correlation of $C$ features and feature $D$. Rules 1-4 are also detected because either total ($C_1-C_2$) or partial of ($C_3-C_4$) have associations with $D$. Rules with features $C_6$ and $C_7$ are not reported because of their zero and negative association~\cite{Antonie2014negative,Li2015associative} with the disease feature. The remaining co-location rules (7-10, 12-14) are detected due to their random correlation with features associated with feature $D$. However, they all have very low $ExpSup(C+D)$ values and can be pruned if a threshold on $ExpSup$ is used as discussed in the experiments with the real data.

\begin{table}[!t]
\renewcommand{\arraystretch}{1.3}
\caption{Co-location rules detected in synthetic data. $ExpSup$ is the value of the expected support of patterns of the form $C+D$, where $C$ is the set of cause features and $D$ is the disease feature}
\label{table_synt}
\centering
\begin{tabular}{|r|l|r|r|}
\hline
N & Co-location Rule & $ExpSup$ & $ExpConf$\\
\hline
1 & $C_1 \to D$ & 763.1 & 0.41\\
\hline
2 & $C_2 \to D$ & 765.8 & 0.42\\
\hline
3 & $C_3 \to D$ & 717.1 & 0.26\\
\hline
4 & $C_4 \to D$ & 807.8 & 0.30\\
\hline
5 & $C_5 \to D$ & 1,256.6 & 0.34\\
\hline
6 & $C_1 + C_2 \to D$ & 432.8 & 0.50\\
\hline
7 & $C_1 + C_4 \to D$ & $1.0 \cdot 10^{-3}$ & 0.82\\
\hline
8 & $C_1 + C_5 \to D$ & 10.6 & 0.49\\
\hline
9 & $C_2 + C_4 \to D$ & 0.4 & 0.49\\
\hline
10 & $C_2 + C_5 \to D$ & 14.4 & 0.44\\
\hline
11 & $C_3 + C_4 \to D$ & 390.5 & 0.53\\
\hline
12 & $ C_5 + C_6 \to D$ & 2.8 & 0.08\\
\hline
13 & $ C_1 + C_2 + C_4 \to D$ & $4.8 \cdot 10^{-4}$ & 0.83\\
\hline
14 & $ C_1 + C_2 + C_5 \to D$ & 7.9 & 0.51\\
\hline
\end{tabular}
\end{table}

The experiment on synthetic data shows that our approach finds co-location rules in which features in the antecedent part are co-located with the feature in the consequent part. A threshold with a relatively low value can help to exclude rules with noise features.

\subsection{Distance between Features}

In this experiment we evaluate the effect of an average distance between features on the expected support. Recall that one of the advantages of our algorithm is that it takes into account a distance between spatial objects, so two objects located close to each other are represented in more transactions than a pair of objects situated farther (Figure~\ref{fig_prox_remote}). Let us consider two scenarios: 1) objects which belong to two distinct features are located on average very close to each other, and 2) they are situated on the farthest possible distance so they are still considered to have neighbourhood relationships. Most previous approaches assign the same prevalence measure value in both cases as long as a neighbourhood relationship is kept. Obviously, it is not correct; the prevalence measure should be higher in the first situation. On the other hand, with our approach in the first case the features are included in more transactions with higher existential probabilities. This leads to a higher prevalence measure than in the second case.

\begin{table}[!t]
\renewcommand{\arraystretch}{1.3}
\caption{The average expected support for ranges of an average distance between two spatial features}
\label{table_synt_remote}
\centering
\begin{tabular}{|r|l|r|}
\hline
N & Range & Average $ExpSup$\\
\hline
1 & [0.0, 0.2) & 1,558.9\\
\hline
2 & [0.2, 0.4) & 1,355.3\\
\hline
3 & [0.4, 0.6) & 1,017.1\\
\hline
4 & [0.6, 0.8) & 649.9\\
\hline
5 & [0.8, 1.0) & 353.3\\
\hline
6 & [1.0, 1.2) & 155.9\\
\hline
7 & [1.2, 1.4) & 52.8\\
\hline
8 & [1.4, 1.6) & 16.6\\
\hline
9 & [1.6, 1.8) & 8.3\\
\hline
10 & [1.8, 2.0) & 5.7\\
\hline
\end{tabular}
\end{table}

For this experiment we create synthetic datasets with two spatial features $f_1$ and $f_2$. The study region is a 100$\times$100 unit square. The buffer size is 1 unit. In each dataset, features have 30 instances each. We randomly place the instances of feature $f_1$ in the study region. One instance of feature $f_2$ is placed on a varying distance $d$ from an instance of $f_1$. The distance $d$ between instances of two features is taken randomly from ten ranges \{[0.0, 0.2), [0.2, 0.4), ..., [1.8, 2.0)\} (given in units). The first range [0.0, 0.2) is for the scenario when features are located very close to each other on average. The last range [1.8, 2.0) simulates a situation when an intersection of each pair of instance buffers is very small. The expected support of pattern $(f_1 \cup f_2)$ is calculated and averaged over 100 synthetic datasets for each of ten ranges.

The results are presented in Table~\ref{table_synt_remote}. As can be observed, the expected support rapidly decreases with the increase in the average distance between instances of features $f_1$ and $f_2$. Expectedly, the range [0.0, 0.2) gets the highest value of the expected support, and the range [1.8, 2.0) has the lowest prevalence value. While a pattern with these features would be considered having the same prevalence measure value in all ten synthetic datasets by most previous algorithms, out transaction-based approach takes into account the actual spatial information and a relative proximity or remoteness of features from each other.

\section{Conclusion}
Co-location pattern and rule mining is one of the tasks of spatial data mining. Discovery of co-location patterns and rules can be useful in many projects and applications and may lead to the discovery of new knowledge in various domains. In this paper we propose a new co-location mining framework which combines classical co-location mining, and uncertain frequent pattern and association rule mining.
The approach was motivated by a real-world application of detecting possible associations of pollutant emission points and childhood cancer cases. We take into account some of the limitations that can prevent previously proposed approaches from being used in some real-world applications and domains. Our novel transactionization method allows the conversion of spatial data into a set of transactions by imposing a regular grid over a given map. Each grid point can be seen as a representation of a study region. Features of objects and their buffers that contain a grid point form a transaction. In addition, our approach takes into account uncertainty of data by storing feature existence probabilities in transactions in order to simulate the real scenarios. A probability of feature presence in a transaction depends on a distance from the feature instance to the respective grid point. A usage of user-defined thresholds on prevalence measures like in previous algorithms is replaced by the statistical test which helps to identify significant co-location patterns and rules that are unlikely to occur only by chance. In order to decrease computation, the filtering techniques are presented which prune candidate patterns and rules that appear to be definitely not significant.

The experiments on two real and synthetic datasets show that our approach finds significant co-location patterns and rules. We deploy three different randomization strategies to find significant co-location rules. The effect of grid granularity is also evaluated. We also compared the results of our method with a baseline method which revealed that our method is not only capable of detecting highly prevalent rules, but is also capable of detecting rules which may not be highly prevalent but statistically significant. A dependence of a prevalence measure value on an average distance between feature instances is shown. The usage of transactions preserves the spatial context and information such as the relative locations of instance objects and distances between them. The consideration of feature presence probabilities helps to distinguish various cases when feature instances are situated at different distances from grid transaction points. We also demonstrate that the difference in the results obtained by our uncertain data model and certain data method can be explained and justified.

The motivating application of this paper has its unique challenges. We examine several factors which affect dispersion of pollutants in the air. In order to more accurately model chemical distribution we used buffer zones differing in their sizes depending on released amounts. Circular buffers are transformed into elliptical figures with the consideration of wind speed and its direction at locations of emitting facilities. Finally, we model uncertainty of a pollutant presence at transaction points. In addition to pollution, other factors can also cause cancer in children. In this paper we do not intend to find true causalities but attempted to identify possible associations of pollutants and childhood cancer. The results that are derived by our algorithm can be useful for domain experts and help in further analysis of pollutant-cancer relationships. These algorithms we propose can also be used to discover co-locations for other diseases and other multiple factors.

\begin{acknowledgements}
This material is supported by Canadian Institutes of Health Research, under grant number REF16629.
\end{acknowledgements}


\end{document}